\documentclass[aps,prl,twocolumn,superscriptaddress]{revtex4-2}

\usepackage{amsmath,amssymb,amsfonts}
\usepackage{accents}
\usepackage{array}
\usepackage[dvipsnames]{xcolor}
\usepackage{hyperref}
\hypersetup{
	pdftex,
	pdfstartview=FitH,
	bookmarksnumbered=true,
	pagebackref,
	colorlinks,
	breaklinks,
	urlcolor=Maroon,
	linkcolor=Maroon,
	citecolor=Bittersweet
}
\usepackage{graphicx}
\DeclareGraphicsExtensions{.pdf,.png,.jpg,.mps}
\usepackage{tikz}
\usetikzlibrary{matrix,arrows.meta,decorations.pathmorphing,decorations.markings,arrows,positioning,intersections,mindmap,backgrounds}
\usepackage{booktabs}
\usepackage{physics}
\usepackage{diagbox}
\usepackage{cleveref}

\usepackage{genyoungtabtikz}
\usepackage{arydshln}

\usepackage{lmodern} 
\usepackage{scalerel,accents,trimclip}
\usepackage{CJK}

\newcommand{\nn}{\nonumber \\}

\newcommand{\newhat}{\scalebox{2}[1.05]{\trimbox{0pt 1.15ex}{\textasciicircum}}}
\newcommand{\mhat}[1]{\accentset{\newhat}{#1}}
\newcommand{\newhatt}{\scalebox{1}[1.0]{\trimbox{0pt 1.1ex}{\textasciicircum}}}
\newcommand{\shat}[1]{\accentset{\newhatt}{#1}}
\newcommand{\sumint}{\scalebox{1.1}{$\displaystyle\int$}\hspace{-4mm}\scalebox{0.9}{$\displaystyle\sum$}}
\newcommand{\thickbar}[1]{\accentset{\raisebox{0.35pt}{\rule{0.4em}{0.05pt}}}{#1}}

\newcommand{\sectionsep}{\vspace{0.5em}}

\newmuskip\pFqmuskip

\newcommand*\pFq[6][8]{
  \begingroup 
  \pFqmuskip=#1mu\relax
  \mathchardef\normalcomma=\mathcode`,
  
  \mathcode`\,=\string"8000
  
  \begingroup\lccode`\~=`\,
  \lowercase{\endgroup\let~}\pFqcomma
  
  {}_{#2}F_{#3}{\left[\genfrac..{0pt}{}{#4}{#5};#6\right]}
  \endgroup
}
\newcommand{\pFqcomma}{{\normalcomma}\mskip\pFqmuskip}

\usepackage{graphicx}

\begin{document}
	
	\begin{CJK*}{UTF8}{gbsn}

	\title{$\rm AdS \times S$ Mellin Bootstrap, Hidden 10d Symmetry and\\
	 Five-point Kaluza-Klein Functions in $\mathcal{N}=4$ SYM}

	\author{Bruno Fernandes}
	\email{cbruno.dfernandes@gmail.com}
	\affiliation{Centro de Fisica do Porto e Departamento de Fisica e Astronomia, Faculdade de Ciencias da Universidade do Porto, Porto 4169-007, Portugal}
	\author{Vasco Gon\c{c}alves}
	\email{vasco.dfg@gmail.com}
	\affiliation{Centro de Fisica do Porto e Departamento de Fisica e Astronomia, Faculdade de Ciencias da Universidade do Porto, Porto 4169-007, Portugal}
	\author{Zhongjie Huang (黄中杰)}
	\email{zjhuang@zju.edu.cn}
	\affiliation{Zhejiang Institute of Modern Physics, School of Physics, Zhejiang University, Hangzhou, Zhejiang 310058, China }
	\affiliation{Joint Center for Quanta-to-Cosmos Physics, Zhejiang University, Hangzhou, Zhejiang 310058, China}
	\author{Yichao Tang (唐一朝)}
	\email{tangyichao@itp.ac.cn}
	\affiliation{Institute of Theoretical Physics, Chinese Academy of Sciences, Beijing 100190, China}
    \affiliation{School of Physical Sciences, University of Chinese Academy of Sciences, Beijing 100049, China}
	\author{Joao Vilas Boas}
	\email{joaomiguelvb@gmail.com}
	\affiliation{School of Mathematical Sciences, Queen Mary University of London, Mile End Road, London, E1 4NS, United Kingdom}
	\author{Ellis Ye Yuan (袁野)}
	\email{eyyuan@zju.edu.cn}
    \affiliation{Zhejiang Institute of Modern Physics, School of Physics, Zhejiang University, Hangzhou, Zhejiang 310058, China }
	\affiliation{Joint Center for Quanta-to-Cosmos Physics, Zhejiang University, Hangzhou, Zhejiang 310058, China}

	\date{\today}

	\begin{abstract}
	    {We propose an AdS$\times$S factorization formula at the level of the generating function for correlators with arbitrary Kaluza-Klein configurations, and implement it in the supergravity limit of $\mathcal{N}=4$ super Yang-Mills. By incorporating this mechanism into Mellin space bootstrap, together with an observed $Z_2$ symmetry under AdS$\ \leftrightarrow\ $S, we manage to simultaneously work out unified formulas both for all five-point half-BPS correlators and for all four-point correlators with one superdescendant. This AdS$\times$S bootstrap method is directly applicable to generic multi-point computation at tree level.}
	\end{abstract}

	\maketitle
	
	\end{CJK*}

\sectionsep\noindent{\bf Introduction.}
Analytic bootstrap has recently boosted the study of four-point correlators of half-BPS operators at various Kaluza-Klein (KK) levels in $\mathcal{N}=4$ super Yang-Mills in the strong coupling region, dual to type IIB superstring in $\mathrm{AdS}_5\times\mathrm{S}^5$ \cite{Rastelli:2016nze,Rastelli:2017udc,Aprile:2017xsp,Alday:2017xua,Alday:2017vkk,Aprile:2017bgs,Aprile:2017qoy,Alday:2019nin,Huang:2021xws,Drummond:2022dxw,Huang:2024rxr,Goncalves:2014ffa,Alday:2018pdi,Alday:2018kkw,Drummond:2019odu,Drummond:2019hel,Drummond:2020dwr,Drummond:2020uni,Aprile:2020mus,Aprile:2022tzr} (see recent reviews \cite{Bissi:2022mrs,Heslop:2022xgp}). This development reveals an interesting connection between interactions among these operators and a hidden 10d conformal symmetry~\cite{Caron-Huot:2018kta}. Correspondingly, a viewpoint has gradually emerged that all four-point correlators involving arbitrary KK configurations can be elegantly packaged into a generating function in position space~\cite{Abl:2020dbx,Caron-Huot:2021usw,Caron-Huot:2023wdh}, or equivalently into an $\mathrm{AdS} \times \mathrm{S}$ Mellin amplitude \cite{Aprile:2020luw,Abl:2020dbx,Aprile:2020mus,Aprile:2022tzr,Huang:2024dxr,Wang:2025pjo} \footnote{See also \cite{Chen:2025yxg,Wu:2025ott} for recent studies on hidden 10d symmetry in giant graviton correlators.}. 

In order to understand whether this hidden higher-dimensional structure is a general feature of the model, it is natural to inspect its potential existence in multi-point correlators. Despite the recent promising progress in a cousin model in $\mathrm{AdS}_5\times\mathrm{S^3}$ \cite{Alday:2022lkk,Alday:2023kfm,Huang:2024dxr,Cao:2023cwa,Cao:2024bky}, the multi-point computation in SYM is still restricted to relatively low KK charges \cite{Goncalves:2019znr,Goncalves:2023oyx,Goncalves:2025jcg}. The main obstacles here are the fast-growing size of the space of internal symmetry structures and of the field contents in the operator product expansion (OPE) due to maximal supersymmetry, which render the original bootstrap strategy impractical. In this work, we overcome these difficulties by formulating a factorization approach in $\mathrm{AdS} \times \mathrm{S}$ Mellin space that simultaneously incorporates all KK contributions. This framework is then integrated into the bootstrap machinery, allowing one to directly construct an ansatz for the $\mathrm{AdS} \times \mathrm{S}$ Mellin amplitudes. The solution follows from this new form of factorization, together with a newly observed $Z_2$ symmetry in $\mathrm{AdS} \leftrightarrow \mathrm{S}$, as well as known constraints from the Drukker-Plefka twist~\cite{Drukker:2009sf,Goncalves:2023oyx,Goncalves:2025jcg}. We demonstrate the effectiveness of this strategy with an explicit five-point computation, in which the generating function for all KK correlators is fully determined.

\sectionsep\noindent{\bf Setup.}
The KK modes we study are the half-BPS operators 
\begin{equation}
    \mathcal{O}_p(X,T) \propto \tr \big[(\phi(X)\cdot T)^p\big] + (\text{multi-trace terms})\,,
\end{equation}
with protected dimension $\Delta=p$. Here we describe the spacetime coordinates by an $\rm SO(4,2)$ null vector $X$ in the embedding space (e.g., \cite{Costa:2011mg}), and the R-symmetry indices in $\phi$ are contracted with an $\rm SO(6)$ null vector $T$, realizing the rank-$p$ symmetric traceless representation. Through AdS/CFT correspondence, they correspond to scalar fields in AdS space at KK level $p$ and are commonly identified as single-particle states of supergravitons \cite{Rastelli:2017udc}. In the large $N$ limit, all multi-trace terms in the operator are suppressed, and we adopt the normalization from \cite{Caron-Huot:2021usw} such that 
\begin{equation}
    \left\langle \mathcal{O}_{p}(X_1, T_1)\mathcal{O}_{p}(X_2, T_2) \right\rangle = \mathcal{N}_{\mathcal{O}} \left(\frac{T_{ij}}{X_{ij}}\right)^p, \ \mathcal{N}_{\mathcal{O}} = \frac{1}{p}\,. 
\end{equation}
Here, $X_{ij} = -X_i \cdot X_j$ and $T_{ij} = T_i \cdot T_j$. 
Under a suitable parametrization, $X_{ij}$ reduces to the ordinary four-dimensional distance $x_{ij}^2$. 

Instead of studying specific KK correlators
\begin{equation}
    G_{\{p_i\}}(X_{ij}, T_{ij}) = \left\langle \mathcal{O}_{p_1}(X_1, T_1)\cdots \mathcal{O}_{p_n}(X_n, T_n) \right\rangle\!
\end{equation}
individually, we can construct a master operator for half-BPS operators by summing over the KK level \cite{Caron-Huot:2021usw}
\begin{equation}
    \mathcal{O}(X,T) = \sum_{p=2}^\infty \mathcal{O}_p(X, T)\,,
\end{equation}
and consider their correlator
\begin{equation}
    \mathcal{G}_n = \langle \mathcal{O}(X_1, T_1)\cdots \mathcal{O}(X_n, T_n) \rangle = \sum_{p_i=2}^\infty G_{\{p_i\}}\, ,
\end{equation}
which acts as a generating function for all KK correlators. This approach offers significant advantage, enabling the simultaneous treatment of all KK levels and revealing previous hidden symmetries. To extract correlators with specific KK levels, we can simply expand the generating function in $T_{ij}$ and pick up terms with the correct weights.

Within the supergravity approximation, the operator spectrum simplifies considerably. Non-protected single-trace operators decouple due to their large anomalous dimensions, thereby streamlining our analysis. Consequently, the OPE of two half-BPS operators contains six types of single-trace operators, as enumerated in Table~\ref{tab:KK_modes}. 

\begin{table}[htbp]
\caption{Quantum numbers of single-trace operators contributing to the OPE of two half-BPS operators.}
\centering
\begin{tabular}{ccccccc}
\toprule
 & $\mathcal{O}_p$ & $\mathcal{J}_p$ & $\mathcal{T}_p$ & $\mathcal{A}_p$ & $\mathcal{C}_p$ & $\mathcal{F}_p$ \\
\midrule
$(\Delta,J)$ & $(p,0)$ & $(p\!+\!1,1)$ & $(p\!+\!2,2)$ & $(p\!+\!2,0)$ & $(p\!+\!3,1)$ & $(p\!+\!4,0)$ \\
\midrule 
$(R,S)$ & $(p,0)$ & $(p\!-\!1,1)$ & $(p\!-\!2,0)$ & $(p\!-\!2,2)$ & $(p\!-\!3,1)$ & $(p\!-\!4,0)$ \\
\bottomrule
\end{tabular}
\label{tab:KK_modes}
\end{table}

These operators are labeled by their conformal dimension $\Delta$ and spin $J$, as well as the ``sphere dimension'' $R$ and ``sphere spin'' $S$, related to the usual $\mathrm{SU}(4)$ Dynkin labels as $[S,R-S,S]$. This terminology will become clear later. We also define the twists $\tau = \Delta - J$ and $\sigma = R + S$ for later convenience. All six operators reside in the same superconformal multiplet $\mathbb O_p$. They are related by the action of supercharges, schematically as 
\begin{align}
    \mathcal{O}_p \xrightarrow{\ Q\thickbar{Q}\ } \mathcal{J}_p \xrightarrow{Q\thickbar{Q}}\,   \begin{array}{c} \mathcal{T}_p \\[1mm] \mathcal{A}_p \end{array}\,  \xrightarrow{Q\thickbar{Q}} \mathcal{C}_p \xrightarrow{Q\thickbar{Q}} \mathcal{F}_p\, .
\end{align}

\sectionsep\noindent{\bf $\rm\bf AdS \times S$ amplitudes.}
The computation of the generating function $\mathcal{G}_n$ is most naturally addressed in $\rm AdS\times S$ Mellin space \cite{Aprile:2020luw}. In $\rm AdS\times S$ Mellin space, we transform both $X_{ij}$ and $T_{ij}$ into the AdS and sphere Mellin variables $\gamma_{ij}$ and $n_{ij}$ \footnote{For simplicity, in the main text we present only the discussion of scalar correlators. The spinning cases can be found in the Supplemental Material.}
\begin{gather}\label{eq:AdSxSG}
    G_{\{p_i\}}(X_{ij},T_{ij}) =\! \sumint' \mathcal{M}_{\{p_i\}}(\gamma_{ij},n_{ij}) \!\prod_{i<j}\! \frac{T_{ij}^{n_{ij}}}{n_{ij}!} \frac{\Gamma(\gamma_{ij})}{X_{ij}^{\gamma_{ij}}}.
\end{gather}
These Mellin variables are symmetric in their two indices and obey the constraints
\begin{align}\label{eq:melincon}
    \forall i:\quad\sum_{j\neq i} \gamma_{ij} = \tau_i=p_i\,,\qquad \sum_{j\neq i} n_{ij} = \sigma_i=p_i\,.
\end{align}
In the sum-integral, we sum over any set of independent $n_{ij}$ at all integers~\footnote{Due to the $1/n_{ij}!$ factors, the sum is automatically truncated to a finite number of integer points such that all $n_{ij}\geq0$.}, and integrate the independent $\gamma_{ij}$ parallel to the imaginary axis:
\begin{align}
    \sumint' = {\sum_{n_{ij}}}' \int_{-i\infty}^{i\infty} {\prod_{\gamma_{ij}}}' \frac{\dd \gamma_{ij}}{2\pi i}\,,
\end{align}
and the prime indicates that $\gamma_{ij}$ and $n_{ij}$ satisfy \eqref{eq:melincon}.

If the amplitude is analytic in $p_i$ (as we always assume in the following), we can further eliminate the explicit $p_i$ dependence by replacing $p_i$ with $n_{ij}$ using \eqref{eq:melincon}, and obtain the $\rm AdS \times S$ amplitude of the generating function
\begin{align}\label{eq:AdSxSGcal}
    \mathcal{G}= \sum_{p_i=2}^{\infty}G_{\{p_i\}} = \sumint\, \mathcal{M}(\gamma_{ij},n_{ij}) \prod_{i<j} \frac{T_{ij}^{n_{ij}}}{n_{ij}!} \frac{\Gamma(\gamma_{ij})}{X_{ij}^{\gamma_{ij}}}\,.
\end{align}
This expression looks very similar to the previous one, but after the sum over $p_i$, the Mellin variables satisfy fewer constraints (hence we drop the prime)
\begin{equation}\label{eq:melincon2}
    \forall i:\quad\sum_{j\neq i} \gamma_{ij} = \sum_{j\neq i} n_{ij}\,.
\end{equation}
These constrain $\gamma_{ij}$, and all $n_{ij}$ are now independent and summed over integers.

In $\rm AdS \times S$ space, the well-known four-point function for arbitrary KK modes can be expressed as \cite{Aprile:2020luw}
\begin{align}
    \mathcal{M}_{4} = \mhat{R}_{1234} \circ \widetilde{\mathcal{M}}_{4}\,\label{eq:fourptRfactor},
\end{align}
where $R_{1234}$ is the superconformal invariant appearing in the superconformal Ward identity \cite{Dolan:2004mu}
\begin{align}
    R_{1234} =& \left(T_{12} T_{34} X_{14} X_{23}\!-\!T_{14} T_{23} X_{12} X_{34}\right)\left(T_{12} T_{34} X_{13} X_{24}\right. \nn 
    &\ \ \left.-T_{13} T_{24} X_{12} X_{34}\right) + (2\leftrightarrow 3)+ (2\leftrightarrow 4)\,,\label{eq:Rfactor4pt}
\end{align}
and $\mhat{R}_{1234}$ denotes its Mellin space counterpart, obtained by translating $X_{ij}$ and $T_{ij}$ into difference operators $\shat{\gamma}_{ij}$ and $\shat{n}_{ij}$
\begin{equation}
\begin{aligned}
    \shat{\gamma}_{ij} \circ F(\gamma_{ij},n_{ij}) &= \gamma_{ij}\, F(\gamma_{ij}+1,n_{ij})\,, \\
    \shat{n}_{ij} \circ F(\gamma_{ij},n_{ij}) &= n_{ij}\, F(\gamma_{ij},n_{ij}-1)\,.
\end{aligned}
\end{equation}    
The reduced amplitude $\widetilde{\mathcal{M}}_{4}$ is a simple rational function
\begin{equation}
    \widetilde{\mathcal{M}}_{4} = \frac{-1}{(\rho_{12}-1)(\rho_{13}-1)(\rho_{14}-1)}\,,
\end{equation}
which depends only on the combination $\rho_{ij} = \gamma_{ij} - n_{ij}$. This is how the hidden 10d symmetry manifests in $\rm AdS \times S$ space: for reduced amplitudes, we can perform the sum over all $n_{ij}$ in \eqref{eq:AdSxSGcal} which leads to an integral over independent $\rho_{ij}$ \cite{Abl:2020dbx}
\begin{equation}
    \hspace{-1mm}\sumint\, \widetilde{\mathcal{M}}(\rho_{ij})\! \prod_{i<j}\! \frac{\Gamma(\gamma_{ij})}{n_{ij}!} \frac{T_{ij}^{n_{ij}}}{X_{ij}^{\gamma_{ij}}}\! =\! \int \widetilde{\mathcal{M}}(\rho_{ij}) \prod_{i<j}\! \frac{\Gamma(\rho_{ij})}{\widetilde{X}_{ij}^{\rho_{ij}}},
\end{equation}
and depends only on the 10d distance $\widetilde{X}_{ij}=X_{ij}-T_{ij}$.

\sectionsep\noindent{\bf Factorization in $\rm\bf AdS \times S$ space.} 
In ordinary Mellin space, factorization of amplitudes \cite{Goncalves:2014rfa} is a powerful tool. The Mellin amplitude $M(\gamma_{ij})$ is a function of AdS variables $\gamma_{ij}$, and the exchange of single-trace operators manifests as a series of poles \footnote{At tree level, multi-trace exchanges  are encoded in poles of $\Gamma$ functions in the conventional Mellin transformation. }. The corresponding residues factorize into products of two sub-amplitudes \cite{Fitzpatrick:2011ia,Goncalves:2014rfa}. For example, when a scalar operator of twist $\tau$ is exchanged, these poles take the form
\begin{align}\label{eq:MellinFac}
    M(\gamma_{ij}) \sim&\ \frac{-2\Gamma(\tau)m!}{(\tau-1)_m}\,  \frac{M_{L,m} \times M_{R,m}}{\gamma_{LR}-\tau-2m}\,,\\
    M_{L,m} =&\ \frac{1}{m!} \left(\mhat{\gamma}_{LL}\right)^m\circ M_{L}(\gamma_{ab})\, ,
\end{align}
and similarly for $M_{R,m}$. Here $(A)_B=\Gamma(A+B)/\Gamma(A)$, $\gamma_{LR} = \sum_{a\in L,b\in R} \gamma_{ab}$, $\mhat{\gamma}_{LL} = \sum_{a,b\in L,a<b} \mhat{\gamma}_{ab}$, $L$ and $R$ denote the sets of external operators in the two sub-amplitudes respectively, and $m=0,1,\dots$ labels the level of conformal descendants.

This factorization property originates from the OPE. It can be derived by solving the Casimir equations in Mellin space 
\begin{align}
    \mhat{C}_2\, M(\gamma_{ij})= \left[ \Delta(\Delta-4)+J(J+2) \right] M(\gamma_{ij})\,,
\end{align}
where $\mhat{C}_2$ denotes the Casimir operator acting on the left-hand side operators. Since this is a group-theoretical result, and the R-symmetry group $\mathrm{SO}(6)$ is simply a Wick rotation of the conformal group $\mathrm{SO}(4,2)$, there exists an $\mathrm{SO}(6)$ analog of the factorization formula for the ``sphere amplitude''. The solutions of the spherical Casimir equations are related to the conformal solutions through
\begin{align}
    (\Delta,J,\tau,m,\gamma_{ij})\leftrightarrow&\ (-R,S,-\sigma,r,-n_{ij})\,, \\ 
    M\big|_{\gamma_{ij}\to \gamma_{ij}\pm 1}\leftrightarrow&\ M\big|_{n_{ij}\to n_{ij}\mp 1}\,,
\end{align}
up to an overall normalization that can be easily determined. This explains our earlier names for $R,S$. 
For an exchange of the sphere scalar $[0,R,0]$, the factorization formula follows directly from \eqref{eq:MellinFac} via Wick rotation
\begin{align}
    M(n_{ab})\Big|_{n_{LR} =\sigma-2r} \!=&\, \frac{(-1)^{r}\,  r!}{\sigma! [\sigma+1]_r } M_{L,r} \times M_{R,r}\,, \\
    M_{L,r} =&\ \frac{1}{r!} \left(\mhat{n}_{LL}\right)^r\circ M_{L}(n_{ab})\,,
\end{align}
with $[A]_B= A!/(A-B)!$.

The Mellin and sphere factorization formulas can be combined into the $\rm AdS\times S$ factorization formula (see the Supplemental Material for details). For a conformal and sphere scalar $\mathcal{X}$ (say $\mathcal O$, $\mathcal F$) 
\begin{align}\label{eq:AdSxSFac}
    \boxed{\mathcal{M}(\gamma_{ij},n_{ij}) \sim\hspace{-2mm} \sum_{m+r=k}\hspace{-2mm} A_{(\tau,m|\sigma,r)}  \frac{\mathcal{M}_{L,m,r} \times  \mathcal{M}_{R,m,r}}{\rho_{LR}-(2\delta+2k)}}\,,
\end{align}
where $2\delta = \tau - \sigma$, and
\begin{align}
    A_{(\tau,m|\sigma,r)} =&\ \frac{1}{\mathcal{N}_\mathcal{X}}\frac{-2\Gamma(\tau)m!}{(\tau-1)_m}\frac{(-1)^{r}\,  r!}{\sigma! [\sigma+1]_r }\,, \\
    \mathcal{M}_{L,m,r} =&\ \frac{\left(\mhat{\gamma}_{LL}\right)^m}{m!}\frac{\left(\mhat{n}_{LL}\right)^r}{r!} \circ \mathcal{M}_{L}(\gamma_{ab},n_{ab})\,,
\end{align}
with $\mathcal{N}_\mathcal{X}$ the normalization factor in the two-point function. One important remark is that $\delta$ does not depend on the KK level. For example, all $\mathcal{O}_p$ satisfy $2\delta=p-p\equiv 0$. Therefore, the entire KK family contributes to the same poles and follows the same factorization formula \eqref{eq:AdSxSFac}. Moreover, substituting $\tau$ and $\sigma$ back with Mellin variables
\begin{align}\label{eq:spurious}
    A_{(\tau,m|\sigma,r)} \Big|_{\substack{\,\tau\to\gamma_{LR}-2m\\ \sigma\to n_{LR}+2r}}
\end{align}
makes \eqref{eq:AdSxSFac} completely independent of KK levels. This allows us to perform a bootstrap analysis for the generating function, and the factorization on \emph{infinitely many} KK modes reduces to the factorization on merely \emph{six} master operators! 

\sectionsep\noindent{\bf $\rm\bf AdS \leftrightarrow S$ $Z_2$ symmetry.}
Before discussing the bootstrap computation, we note a remarkable $Z_2$ symmetry that interchanges variables in AdS and S. This symmetry manifests in the kinematic variables in the embedding formalism ($X$ and $T$) and the $\rm AdS\times S$ formalism ($\gamma$ and $n$). Moreover, it appears in the spectrum of single-trace operators, where taking 
\begin{align}
    (\Delta,J) \leftrightarrow (-R,S),\quad p \leftrightarrow -p
\end{align}
in Table \ref{tab:KK_modes} maps each single-trace operator to itself, except $\mathcal{T}_p$ and $\mathcal{A}_p$ are interchanged. This symmetry also holds at the correlator level \footnote{When limited on the reduced correlators, this $Z_2$ symmetry can be viewed as an element in the hidden 10d conformal group. }
\begin{align}
    \mathcal{M}_4(\gamma_{ij},n_{ij}) = \mathcal{M}_4(-n_{ij},-\gamma_{ij})\,.
\end{align}
Since it exchanges $p \leftrightarrow -p$, the $Z_2$ symmetry is invisible for a particular KK correlator, but observable only when considering the entire family or the generating function of the correlators \footnote{The same $Z_2$ symmetry was discovered in \cite{Aprile:2020mus} for stringy corrections. We thank Francesco Aprile and Michele Santagata for pointing out this.}.

More importantly, the symmetry persists for correlators involving superdescendants $\mathcal{X}=\mathcal{J},\mathcal{T},\mathcal{A},\mathcal{C},\mathcal{F}$. This can be proven using the superspace method, where superdescendants are extracted from the superoperator $\mathbb{O}_p$ via a differential operator
\begin{equation}
    \mathcal{X}_p = \mathcal{D}^{(\mathcal{X}_p)} \mathbb{O}_p \big|_{\varrho=\lambda=0}\,,
\end{equation}
and the operator $\mathcal{D}^{(\mathcal{X}_p)}$ turns out to be $Z_2$-symmetric (see Supplemental Material for details). It is worth noting that, since both the $\rm AdS \times S$ factorization formula and all sub-amplitudes (including those with superdescendants) respect the $Z_2$ symmetry, for arbitrary $n$-point functions we expect to have 
\begin{align}
    \mathcal{M}_n(\gamma_{ij},n_{ij}) = (-)^n \mathcal{M}_n(-n_{ij},-\gamma_{ij})\,,
\end{align}
where $(-)^n$ comes from the normalization $\mathcal{N}_\mathcal{O}$.

\sectionsep\noindent{\bf Four-point function.} 
We first apply the $\rm AdS\times S$ factorization formula on the four-point amplitude $\mathcal{M}_4$ and see what we can learn from the computation. In the s-channel, $\rho_{LR} = -2\rho_{12}$, and we have the following poles 
\begin{align}
    \mathcal{M}_{4} \sim   \frac{\mathcal{Q}_0(\gamma_{ij},n_{ij})}{\rho _{12}} + \frac{\mathcal{Q}_1(\gamma_{ij},n_{ij})}{\rho _{12}+1},
\end{align}
where $\mathcal{Q}_0$ and $\mathcal{Q}_1$ can be obtained from \eqref{eq:fourptRfactor}. Both $\mathcal{O}$ and $\mathcal{J}$ have $\delta=0$ and contribute to the leading residue $\mathcal{Q}_0$, and we can extract the sub-amplitudes $\mathcal{M}_{3,\mathcal{O}}$, $\mathcal{M}_{3,\mathcal{J}}$ of correlators $\langle \mathcal{O}\mathcal{O}\mathcal{O} \rangle$, $\langle \mathcal{J}\mathcal{O}\mathcal{O} \rangle$ from it. For instance,
\begin{align}
    \mathcal{M}_{3,\mathcal{O}} =&\ n_{12}n_{1I}n_{2I},
\end{align}
where $I$ labels the exchanged operator \footnote{The amplitude $\mathcal{M}_{3,\mathcal{O}}$ might appear to violate the $Z_2$ symmetry $\gamma_{ij}\leftrightarrow -n_{ij}$, but in this three-point amplitude we have $\gamma_{ij}\equiv n_{ij}$. }. One can recover the well-known three-point function result by substituting $\mathcal{M}_{3,\mathcal{O}}$ back in \eqref{eq:AdSxSG}. 
Note that $\mathcal M_4$ only contains two poles $\rho_{12}=0,-1$, but the gluing of the three-point amplitudes $\mathcal{M}_{3,\mathcal{X}}$ according to \eqref{eq:AdSxSFac} produces higher-descendant poles $\rho_{12}=-k$ for $k\geq2$. By requiring the residue at $k=1$ matches $\mathcal Q_1$ and the residues for all $k\geq2$ vanish after summing over the contribution from all six master operators, we extract all other superdescendant amplitudes $\mathcal{M}_{3,\mathcal{X}}$ with $\mathcal{X}=\mathcal{T},\mathcal{A},\mathcal{C},\mathcal{F}$ (which are recorded in the Supplemental Material). The delicate cancellation of the higher-descendant poles between different superdescendants can be regarded as a consequence of supersymmetry. Moreover, the coefficient \eqref{eq:spurious} introduces spurious poles of $\gamma_{LR}$ and $n_{LR}$ (instead of $\rho_{LR}$) for individual master exchanges. In principle, these spurious poles could spoil the hidden 10d symmetry in $\mathcal{M}_4$. However, in the end they all cancel and result in a rational function with only $\rho$ poles. We can say, in a rough sense, that supersymmetry leads to hidden 10d symmetry in $\mathcal{M}_4$. In the following bootstrap of five-point functions, we will flip the logic around, using hidden 10d symmetry as an input to bootstrap $\mathcal M_5$, as well as all the unknown four-point sub-amplitudes involving superdescendants.

\sectionsep\noindent{\bf Bootstrapping five-point function.} 
Now we bootstrap the five-point $\rm AdS\times S$ amplitude $\mathcal{M}_5$, together with all four-point superdescendant amplitudes $\mathcal{M}_{4,\mathcal{X}}$ (the amplitude of $\langle \mathcal{X}\mathcal{O}\mathcal{O}\mathcal{O} \rangle$, $\mathcal{X}=\mathcal{J},\mathcal{T},\mathcal{A},\mathcal{C},\mathcal{F}$), in \emph{one bootstrap computation}. The strategy is as follows: 
\\[1mm]
\noindent{\bf (I) Hidden 10d symmetry:} We first write down an ansatz satisfying (a weaker version of) hidden 10d symmetry, in the sense that there are only truncated $\rho$ poles in the amplitudes~\footnote{We will see that, after the bootstrap computation, the result can indeed be put in a factorized form $\unexpanded{\mathcal{M}=\sum\mhat{R}\circ\widetilde{\mathcal M}}$ with superconformal invariants $R$ and reduced amplitudes $\unexpanded{\widetilde{\mathcal{M}}}$ depending only on $\rho$.}
\begin{align}
    \mathcal{M}_5 =&\! \sum_{k_1,k_2=0} \frac{P^{(k_1,k_2)} (\gamma_{ij},n_{ij})}{(\rho_{12}\!+\!k_1)(\rho_{45}\!+\!k_2)} \!+\! \sum_{k_3=0} \frac{P^{(k_3)}(\gamma_{ij},n_{ij})}{\rho_{12}+k_3}  \nn
    & + P(n_{ij}) + (\text{perms}), \\[2mm]
    \mathcal{M}_{4,\mathcal{X}} \!=& \!\sum_{k_4=0}^{\delta+1}\!\! \frac{Q^{(k_4)}_{\mathcal{X}}(\gamma_{ij},n_{ij})}{\rho_{12}+k_4} \!+\! Q_{\mathcal{X}}(\gamma_{ij},n_{ij}) \!+\! (\text{perms}), 
\end{align}
where $P$ and $Q$'s are polynomials of $\gamma_{ij}$ and $n_{ij}$. When $\mathcal{X}$ has nonzero spin, $\mathcal{M}_{4,\mathcal{X}}$ will have spinning components, but we suppress their labels above. These components are not independent and satisfy a ``gauge invariance'' condition \cite{Goncalves:2014rfa} (see Supplemental Material for details). 

In the ansatz, $P^{(k_1,k_2)}$ and $Q^{(k_4)}_{\mathcal{X}}$ depend quadratically on $\gamma$, while $P^{(k_3)}$ and $Q_{\mathcal{X}}$ are linear in $\gamma$ and $P$ indepdendent of $\gamma$. This power counting comes from the flat space limit in conventional Mellin amplitudes \cite{Goncalves:2023oyx}. For the total power of $\gamma$ and $n$, the amplitudes should scale like $\mathcal{M}_5\sim \lambda^7$ and $\mathcal{M}_{4,\mathcal{X}} \sim \lambda^{5,6,6,8,9}$ for $\mathcal{X}=\mathcal{J},\mathcal{T},\mathcal{A},\mathcal{C},\mathcal{F}$ when $\gamma,n\sim \lambda\to\infty$. The power counting for $\mathcal{M}_5$ is based on the factorization of $\mathcal O$ in $\mathcal{M}_5 \supset \mathcal{M}_{3,\mathcal{O}}\times \mathcal{M}_{4,\mathcal{O}}/\rho_{LR} \sim \lambda^{3+5-1}$, and those for $\mathcal{M}_{4,\mathcal{X}}$ are to ensure all these sub-amplitudes give rise to $\lambda^7$ in $\mathcal{M}_5$ using the factorization formula. 

\begin{figure}[h]
    \centering
    \begin{tikzpicture}[scale=0.6]
        \begin{scope}
            \draw[thick, arrows = {-Stealth[reversed, reversed]}] (-0.5,0) -- (5,0);
            \draw[thick, arrows = {-Stealth[reversed, reversed]}] (0,-0.5) -- (0,5); 
            \draw[dotted, step=1] (-0.3,-0.3) grid (4.9,4.9); 
           \draw[dashed] (0,1) -- (3.6,4.6);
            \foreach \x in {0,...,4} \foreach \y in {0,...,\x} \draw[gray, thick, fill=white] (\x,\y) circle [radius=0.09];
            \foreach \x in {0,...,3} \draw[gray, thick, fill=white] (\x,\x+1) circle [radius=0.09];
            \draw[BrickRed, very thick, fill] (0,0) circle [radius=0.09] (0,1) circle [radius=0.09] (1,0) circle [radius=0.09] (1,1) circle [radius=0.09] (1,2) circle [radius=0.09] (2,1) circle [radius=0.09]; 
            \node[anchor=north] at (5,0) {$k_1$};
            \node[anchor=east] at (0,5) {$k_2$}; 
        \end{scope}
        \begin{scope}[xshift=6.5cm, yshift=1.3cm]
            \draw[thick, arrows = {-Stealth[reversed, reversed]}] (-0.5,0) -- (5,0);
            \draw[dotted, thick] (0,-0.5) -- (0,0.5);
            \node[anchor=north] at (0,-0.5) {$0$};
            \foreach \x in {0,...,4} \draw[gray, thick, fill=white] (\x,0) circle [radius=0.09];
            \foreach \x in {0,1,2} \draw[BrickRed, very thick, fill] (\x,0) circle [radius=0.09];
            \node[anchor=north] at (5,0) {$k_3$};
        \end{scope}
        \begin{scope}[yshift=0.5cm]
            \draw[gray, thick, fill=white] (6,4) circle [radius=0.09];
            \node[anchor=west, align=left] at (6,4) {\scriptsize : possible position of residues};
            \draw[BrickRed, very thick, fill] (6,3.5) circle [radius=0.09];
            \node[anchor=west] at (6,3.5) {\scriptsize : non-zero residues};
        \end{scope}
    \end{tikzpicture} 
    \caption{The position of non-zero residues in the five-point amplitude ansatz. }
    \label{fig:residues}
\end{figure}
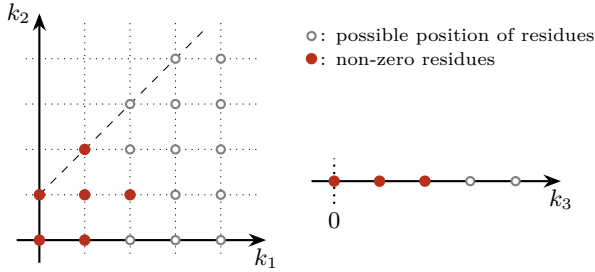

The truncation of $\rho$ poles in $\mathcal{M}_{4,\mathcal{X}}$ follows also from considerations of the factorization on $\mathcal{O}$. When gluing $\mathcal{M}_{3,\mathcal{O}}$ and $\mathcal{M}_{4,\mathcal{O}}$ in the channel (12)-(345), poles in $\mathcal{M}_{4,\mathcal{O}}$ give rise to double poles in $\mathcal{M}_5$ (see \eqref{eq:5ptfac1}), located at the spots in Figure \ref{fig:residues}, inside the region under the dashed line $k_2 = k_1 +1$. The truncation of poles in $\mathcal{M}_{4,\mathcal{X}}$ ensures that double poles due to exchange of $\mathcal{X}$ are also located inside the same region. We do not 
truncate poles in the ansatz of $\mathcal{M}_5$ by hand~\footnote{In fact, we did not write down an ansatz for $P^{(k_1,k_2)}$ and $P^{(k_3)}$ explicitly in the actual bootstrap computation. All constraints on these polynomials should be understood as imposed directly on the RHS of \eqref{eq:5ptfac1} and \eqref{eq:5ptfac2}.}, but they turn out to be truncated automatically after the bootstrap computation.

In principle, one can write down a larger ansatz with higher-degree polynomials and more poles. The bootstrap computation justifies our minimal ansatz above, which already yields a consistent solution that passes various checks. 
\\[1mm]
\noindent{\bf (II) Crossing symmetry:} The ansatz should satisfy crossing symmetry. This means that $\mathcal{M}_5$ should be invariant under $S_5$, and $\mathcal{M}_{4,\mathcal{X}}$ should be invariant under the $S_3$ symmetry among the three superprimaries $\mathcal{O}$. This is largely taken care of by the ``+ (\text{perms})'' term in the ansatz, but still imposes some additional constraints on the polynomials when it maps the poles into the same channel. For example, $(1,2)\leftrightarrow (4,5)$ leaves the channel $\rho_{12}\rho_{45}$ unchanged, and requires that
\begin{align}
    P^{(k_1,k_2)}(\gamma_{ij},n_{ij}) = P^{(k_2,k_1)}(\gamma_{ij},n_{ij})\big|_{(1,2)\leftrightarrow (4,5)}.
\end{align}
There are similar constraints on other $P$ and $Q$'s, but we refrain to write them down explicitly. 
\\[1mm]
\noindent{\bf (III) $Z_2$ symmetry:} In addition, we impose the $\rm AdS \leftrightarrow S$ symmetry on $ \mathcal{M}_{4,\mathcal{X}}$. It takes the following form on the scalar operator $\mathcal{F}$
\begin{align}
    \mathcal{M}_{4,\mathcal{F}}(\gamma_{ij},n_{ij}) =&\ \mathcal{M}_{4,\mathcal{F}}(-n_{ij},-\gamma_{ij}),
\end{align}
and those with spinning operators are presented in the Supplemental Material. This $Z_2$ symmetry halves the unknown coefficients in the ansatz of $\mathcal M_{4,\mathcal J/\mathcal C/\mathcal F}$, and reduces the ansatz of $\mathcal{M}_{4,\mathcal{A}}$ to that of $\mathcal{M}_{4,\mathcal{T}}$ completely. 
\\[1mm]
\noindent{\bf (IV) $\rm\bf AdS\times S$ factorization:} Using the $\rm AdS \times S$ factorization formula, we can relate the polynomials in $\mathcal{M}_5$ and $\mathcal{M}_{4,\mathcal{X}}$ by
\begin{align}
    P^{(k_1,k_2)} =&\ \sum_{\mathcal{X}} \text{Fac}_{\mathcal{X},k_1}\left[\mathcal{M}_{3,\mathcal{X}} \times \widetilde{Q}^{(k_2)}_{\mathcal{X}}\right], \label{eq:5ptfac1}\\
    P^{(k_3)} =&\ \sum_{\mathcal{X}}\text{Fac}_{\mathcal{X},k_3}\left[\mathcal{M}_{3,\mathcal{X}} \times \widetilde{Q}_{\mathcal{X}}\right], \label{eq:5ptfac2}
\end{align}
where $\text{Fac}_{\mathcal{X},k}$ stands for the factorization formula for the residue of $\mathcal{X}$-exchange at pole $\rho_{12}=-k$. $\mathcal{X}$ runs over all six master operators, and $\widetilde{Q} \equiv Q|_{(1,2,3,4)\to (I,3,4,5)}$ is the relabeling of points in the sub-amplitudes. Since $P$'s are polynomials, these two equations implicitly require that there are no spurious poles on the RHS after applying the factorization formula. The absence of spurious poles turns out to be a very strong constraint, which solves a large part of unknowns at this stage. 

It is worth noting that, after imposing the factorization formula, all the superdescendant amplitudes $\mathcal{M}_{4,\mathcal{X}}$ are completely fixed, without even using supersymmetry! Also, the poles in $\mathcal{M}_5$ turns out to be truncated at $k\leq 2$ (we checked the result up to $k=5$). The non-zero residues are marked red in Figure \ref{fig:residues}.  
\\[1mm]
\noindent{\bf (V) Drukker-Plefka twist:} There remains a regular term $P(n_{ij})$ to be determined in $\mathcal{M}_5$. This term is not probed by the factorization, and we fix it by the Drukker-Plefka (DP) twist \cite{Drukker:2009sf} (note that this twist also respect $Z_2$ symmetry) 
\begin{align}
    G_{\{p_i\}}(X_{ij},T_{ij})\big|_{T_{ij}\to X_{ij}} = \text{const.}
\end{align}
In $\rm AdS\times S$ space, we can impose this condition by first decompose $\mathcal M_5(\rho_{ij}+n_{ij},n_{ij})$ (which is a polynomial with respect to $n_{ij}$) in the basis of factorial powers $[n_{ij}]_{c_{ij,k}}$
\begin{align}
    \mathcal{M}(\rho_{ij}+n_{ij},n_{ij}) \equiv \sum_{k} \mathcal{M}^k(\rho_{ij}) \prod_{i<j}\,[n_{ij}]_{c_{ij,k}}\,
\end{align}
and then require the function to vanish upon replacing
\begin{align} 
    [n_{ij}]_{c_{ij,k}} \to \left(\frac{a_i a_j}{1-a_ia_j}\right)^{c_{ij,k}} (\rho_{ij})_{c_{ij,k}},
\end{align}
with $a_i$ some auxiliary variables (see Supplemental Material for the derivation). Imposing the $\rm AdS\times S$ DP twist completely determines $\mathcal{M}_5$.

\sectionsep\noindent{\bf Result.}
The full expression for $\mathcal{M}_5$ is too lengthy to include in the main text. We therefore present selected components in the Supplemental Material and provide the complete result as an auxiliary file. Our results pass several non-trivial consistency checks: $\mathcal{M}_5$ reduces to the known result $M_{pp222}$ in \cite{Goncalves:2023oyx}, and the amplitudes $\mathcal{M}_{4,\mathcal{X}}$ for $\mathcal{X} = \mathcal{J},\mathcal{T},\mathcal{A}$ agree with results independently derived using the superspace method (see Supplemental Material). 

Furthermore, we confirm that $\mathcal{M}_5$ can be recast in a factorized form that generalizes \eqref{eq:fourptRfactor} to five points
\begin{align}\label{eq:M5R}
    \mathcal{M}_{5}(\gamma_{ij},n_{ij}) = \sum_k \mhat{R}^{(k)} \circ \widetilde{\mathcal{M}}^{(k)}_{5}(\rho_{ij}) + (\text{perms}),
\end{align}
where $\widetilde{\mathcal{M}}^{(k)}_{5}$ are simple rational functions with five distinct poles 
\begin{align}
    \widetilde{\mathcal{M}}^{(k)}_{5}(\rho_{ij}) =\! \prod_{\text{(\# of poles)}=5} \ \frac{1}{\rho_{ij}-1}\,,
\end{align}
and $R^{(k)}$ are (the bosonic component of) some superconformal invariants \footnote{See \cite{Heslop:2022xgp} and Supplemental Material for the way to construct these superconformal invariants. }, manifesting supersymmetry in $\mathcal{M}_{5}$. Notably, we did not impose supersymmetry as a bootstrap constraint (except for the DP twist), but it emerges as a result from hidden 10d symmetry. However, the representation \eqref{eq:M5R} we found is not unique. The ambiguities arise from certain superconformal invariants that sum up to 0 in \eqref{eq:M5R}. Note that, some of these zero functions in Mellin space are actually nonzero in position space due to the contour pinching issue \cite{Rastelli:2017udc}, and the corresponding ambiguities can be fixed in position space. We leave it for future work. 


\sectionsep\noindent{\bf Outlook.}
The obvious next target is computing complete six- and higher-point amplitudes for half-BPS operators, where our method directly applies and gives all KK correlators at given points once and for all. In principle, this method can be applied recursively: we expect that bootstrapping $\mathcal{M}_{n}$ would generate $\mathcal{M}_{n-1,\mathcal{X}}$ as byproducts, and these $\mathcal{M}_{n-1,\mathcal{X}}$ can then be fed into the factorization constraints of $\mathcal{M}_{n+1}$
\begin{equation}
    \mathcal{M}_{n+1} \sim \sum_{\mathcal{X}}\mathcal{M}_{4,\mathcal{X}} \times \mathcal{M}_{n-1,\mathcal{X}} \sim \sum_{\mathcal{X}}\mathcal{M}_{3,\mathcal{X}} \times \mathcal{M}_{n,\mathcal{X}}
\end{equation}
to bootstrap $\mathcal{M}_{n+1}$ itself as well as $\mathcal{M}_{n,\mathcal{X}}$. Another exciting direction is to further incorporate supersymmetry in this formalism and combine the six master operators into a super master operator $\mathbb{O}$, as was studied in \cite{Caron-Huot:2023wdh} at weak coupling. This operator would be the only object we need to consider in the factorization of tree-level functions, and may help to reveal the relation between supersymmetry and hidden 10d symmetry in $\mathcal{N}=4$ SYM at strong coupling. 

Multi-trace operators naturally appear when we consider the OPE in higher-point correlators. However, very little is known about the spectrum and correlators of these multi-trace operators (see \cite{Ceplak:2021wzz,Ma:2022ihn,Fardelli:2024heb,Aprile:2024lwy,Kravchuk:2024wmv,Bissi:2024tqf,Aprile:2025hlt} for some recent progress). By analysing the OPE data in our five-point function, we can extract the triple-trace anomalous dimensions for arbitrary KK modes, which would be crucial to the study of multi-particle bound states in AdS, and might reveal how hidden 10d symmetry encodes in multi-trace spectrum.

So far, the vast majority of results for more than four points have been limited to the supergravity approximation, thus it would be interesting to study stringy and loop corrections to our result~\cite{VilasBoas:2025vvw}, and also investigate at the weak coupling region \cite{Bargheer:2022sfd,Bargheer:2025uai}. This could give us hints of different representations for the amplitude. 

A key ingredient in establishing the double-copy structure of four-point amplitudes in AdS~\cite{Zhou:2021gnu} was the identification of a universal supersymmetric prefactor. In this work, we proposed an extension of this prefactor to five-point functions. It would be interesting to explore whether a similar double-copy structure exists at five points, both in the half-BPS sector studied here and for gluon correlators in models such as $\rm AdS_5\times S^3$ \cite{Huang:2024dxr}.    

\sectionsep  
\begin{acknowledgments}
    
	\noindent{\bf Acknowledgments.}
	The authors would like to thank Ant\'onio Antunes, Francesco Aprile, Paul Heslop, Xiang Li, Rodolfo Russo, Congkao Wen, Jiarong Zhang and Xinan Zhou for useful discussions. ZH and EYY are supported by National Science Foundation of China under Grant No.~12175197 and Grand No.~12347103. EYY is also supported by National Science Foundation of China under Grant No.~11935013, and by the Fundamental Research Funds for the Chinese Central Universities under Grant No.~226-2022-00216. Centro de Fisica do Porto is partially funded by Fundacao para a Ciencia e a Tecnologia (FCT) under the grant UID04650-FCUP. V.G. is supported by Fundacao para a Ciencia e Tecnologia (FCT) under the grant CEECIND/03356/2022, by FCT grant 2024.00230.CERN and HORIZON-MSCA-2023-SE-01-101182937-HeI. B.F. is supported by Funda\c{c}\~{a}o para a Ci\^{e}ncia e a Tecnologia, under the IDPASC doctoral program, under Grant No. PRT/BD/154692/2022, by FCT grant 2024.00230.CERN and HORIZON-MSCA-2023-SE-01-101182937-HeI. JVB is supported by the UK EPSRC grant 'CFT and Gravity: Heavy States and Black Holes' EP/W019663/1. No new data were generated or analysed during this study.
\end{acknowledgments}

\bibliography{refs}

\begin{thebibliography}{70}%
\makeatletter
\providecommand \@ifxundefined [1]{%
 \@ifx{#1\undefined}
}%
\providecommand \@ifnum [1]{%
 \ifnum #1\expandafter \@firstoftwo
 \else \expandafter \@secondoftwo
 \fi
}%
\providecommand \@ifx [1]{%
 \ifx #1\expandafter \@firstoftwo
 \else \expandafter \@secondoftwo
 \fi
}%
\providecommand \natexlab [1]{#1}%
\providecommand \enquote  [1]{``#1''}%
\providecommand \bibnamefont  [1]{#1}%
\providecommand \bibfnamefont [1]{#1}%
\providecommand \citenamefont [1]{#1}%
\providecommand \href@noop [0]{\@secondoftwo}%
\providecommand \href [0]{\begingroup \@sanitize@url \@href}%
\providecommand \@href[1]{\@@startlink{#1}\@@href}%
\providecommand \@@href[1]{\endgroup#1\@@endlink}%
\providecommand \@sanitize@url [0]{\catcode `\\12\catcode `\$12\catcode
  `\&12\catcode `\#12\catcode `\^12\catcode `\_12\catcode `\%12\relax}%
\providecommand \@@startlink[1]{}%
\providecommand \@@endlink[0]{}%
\providecommand \url  [0]{\begingroup\@sanitize@url \@url }%
\providecommand \@url [1]{\endgroup\@href {#1}{\urlprefix }}%
\providecommand \urlprefix  [0]{URL }%
\providecommand \Eprint [0]{\href }%
\providecommand \doibase [0]{https://doi.org/}%
\providecommand \selectlanguage [0]{\@gobble}%
\providecommand \bibinfo  [0]{\@secondoftwo}%
\providecommand \bibfield  [0]{\@secondoftwo}%
\providecommand \translation [1]{[#1]}%
\providecommand \BibitemOpen [0]{}%
\providecommand \bibitemStop [0]{}%
\providecommand \bibitemNoStop [0]{.\EOS\space}%
\providecommand \EOS [0]{\spacefactor3000\relax}%
\providecommand \BibitemShut  [1]{\csname bibitem#1\endcsname}%
\let\auto@bib@innerbib\@empty
\bibitem [{\citenamefont {Rastelli}\ and\ \citenamefont
  {Zhou}(2017)}]{Rastelli:2016nze}%
  \BibitemOpen
  \bibfield  {author} {\bibinfo {author} {\bibfnamefont {L.}~\bibnamefont
  {Rastelli}}\ and\ \bibinfo {author} {\bibfnamefont {X.}~\bibnamefont
  {Zhou}},\ }\bibfield  {title} {\bibinfo {title} {{Mellin amplitudes for
  $AdS_5\times S^5$}},\ }\href {https://doi.org/10.1103/PhysRevLett.118.091602}
  {\bibfield  {journal} {\bibinfo  {journal} {Phys. Rev. Lett.}\ }\textbf
  {\bibinfo {volume} {118}},\ \bibinfo {pages} {091602} (\bibinfo {year}
  {2017})},\ \Eprint {https://arxiv.org/abs/1608.06624} {arXiv:1608.06624
  [hep-th]} \BibitemShut {NoStop}%
\bibitem [{\citenamefont {Rastelli}\ and\ \citenamefont
  {Zhou}(2018)}]{Rastelli:2017udc}%
  \BibitemOpen
  \bibfield  {author} {\bibinfo {author} {\bibfnamefont {L.}~\bibnamefont
  {Rastelli}}\ and\ \bibinfo {author} {\bibfnamefont {X.}~\bibnamefont
  {Zhou}},\ }\bibfield  {title} {\bibinfo {title} {{How to Succeed at
  Holographic Correlators Without Really Trying}},\ }\href
  {https://doi.org/10.1007/JHEP04(2018)014} {\bibfield  {journal} {\bibinfo
  {journal} {JHEP}\ }\textbf {\bibinfo {volume} {04}},\ \bibinfo {pages}
  {014}},\ \Eprint {https://arxiv.org/abs/1710.05923} {arXiv:1710.05923
  [hep-th]} \BibitemShut {NoStop}%
\bibitem [{\citenamefont {Aprile}\ \emph
  {et~al.}(2018{\natexlab{a}})\citenamefont {Aprile}, \citenamefont {Drummond},
  \citenamefont {Heslop},\ and\ \citenamefont {Paul}}]{Aprile:2017xsp}%
  \BibitemOpen
  \bibfield  {author} {\bibinfo {author} {\bibfnamefont {F.}~\bibnamefont
  {Aprile}}, \bibinfo {author} {\bibfnamefont {J.~M.}\ \bibnamefont
  {Drummond}}, \bibinfo {author} {\bibfnamefont {P.}~\bibnamefont {Heslop}},\
  and\ \bibinfo {author} {\bibfnamefont {H.}~\bibnamefont {Paul}},\ }\bibfield
  {title} {\bibinfo {title} {{Unmixing Supergravity}},\ }\href
  {https://doi.org/10.1007/JHEP02(2018)133} {\bibfield  {journal} {\bibinfo
  {journal} {JHEP}\ }\textbf {\bibinfo {volume} {02}},\ \bibinfo {pages}
  {133}},\ \Eprint {https://arxiv.org/abs/1706.08456} {arXiv:1706.08456
  [hep-th]} \BibitemShut {NoStop}%
\bibitem [{\citenamefont {Alday}\ and\ \citenamefont
  {Bissi}(2017)}]{Alday:2017xua}%
  \BibitemOpen
  \bibfield  {author} {\bibinfo {author} {\bibfnamefont {L.~F.}\ \bibnamefont
  {Alday}}\ and\ \bibinfo {author} {\bibfnamefont {A.}~\bibnamefont {Bissi}},\
  }\bibfield  {title} {\bibinfo {title} {{Loop Corrections to Supergravity on
  $AdS_5 \times S^5$}},\ }\href
  {https://doi.org/10.1103/PhysRevLett.119.171601} {\bibfield  {journal}
  {\bibinfo  {journal} {Phys. Rev. Lett.}\ }\textbf {\bibinfo {volume} {119}},\
  \bibinfo {pages} {171601} (\bibinfo {year} {2017})},\ \Eprint
  {https://arxiv.org/abs/1706.02388} {arXiv:1706.02388 [hep-th]} \BibitemShut
  {NoStop}%
\bibitem [{\citenamefont {Alday}\ and\ \citenamefont
  {Caron-Huot}(2018)}]{Alday:2017vkk}%
  \BibitemOpen
  \bibfield  {author} {\bibinfo {author} {\bibfnamefont {L.~F.}\ \bibnamefont
  {Alday}}\ and\ \bibinfo {author} {\bibfnamefont {S.}~\bibnamefont
  {Caron-Huot}},\ }\bibfield  {title} {\bibinfo {title} {{Gravitational
  S-matrix from CFT dispersion relations}},\ }\href
  {https://doi.org/10.1007/JHEP12(2018)017} {\bibfield  {journal} {\bibinfo
  {journal} {JHEP}\ }\textbf {\bibinfo {volume} {12}},\ \bibinfo {pages}
  {017}},\ \Eprint {https://arxiv.org/abs/1711.02031} {arXiv:1711.02031
  [hep-th]} \BibitemShut {NoStop}%
\bibitem [{\citenamefont {Aprile}\ \emph
  {et~al.}(2018{\natexlab{b}})\citenamefont {Aprile}, \citenamefont {Drummond},
  \citenamefont {Heslop},\ and\ \citenamefont {Paul}}]{Aprile:2017bgs}%
  \BibitemOpen
  \bibfield  {author} {\bibinfo {author} {\bibfnamefont {F.}~\bibnamefont
  {Aprile}}, \bibinfo {author} {\bibfnamefont {J.~M.}\ \bibnamefont
  {Drummond}}, \bibinfo {author} {\bibfnamefont {P.}~\bibnamefont {Heslop}},\
  and\ \bibinfo {author} {\bibfnamefont {H.}~\bibnamefont {Paul}},\ }\bibfield
  {title} {\bibinfo {title} {{Quantum Gravity from Conformal Field Theory}},\
  }\href {https://doi.org/10.1007/JHEP01(2018)035} {\bibfield  {journal}
  {\bibinfo  {journal} {JHEP}\ }\textbf {\bibinfo {volume} {01}},\ \bibinfo
  {pages} {035}},\ \Eprint {https://arxiv.org/abs/1706.02822} {arXiv:1706.02822
  [hep-th]} \BibitemShut {NoStop}%
\bibitem [{\citenamefont {Aprile}\ \emph
  {et~al.}(2018{\natexlab{c}})\citenamefont {Aprile}, \citenamefont {Drummond},
  \citenamefont {Heslop},\ and\ \citenamefont {Paul}}]{Aprile:2017qoy}%
  \BibitemOpen
  \bibfield  {author} {\bibinfo {author} {\bibfnamefont {F.}~\bibnamefont
  {Aprile}}, \bibinfo {author} {\bibfnamefont {J.~M.}\ \bibnamefont
  {Drummond}}, \bibinfo {author} {\bibfnamefont {P.}~\bibnamefont {Heslop}},\
  and\ \bibinfo {author} {\bibfnamefont {H.}~\bibnamefont {Paul}},\ }\bibfield
  {title} {\bibinfo {title} {{Loop corrections for Kaluza-Klein AdS
  amplitudes}},\ }\href {https://doi.org/10.1007/JHEP05(2018)056} {\bibfield
  {journal} {\bibinfo  {journal} {JHEP}\ }\textbf {\bibinfo {volume} {05}},\
  \bibinfo {pages} {056}},\ \Eprint {https://arxiv.org/abs/1711.03903}
  {arXiv:1711.03903 [hep-th]} \BibitemShut {NoStop}%
\bibitem [{\citenamefont {Alday}\ and\ \citenamefont
  {Zhou}(2020)}]{Alday:2019nin}%
  \BibitemOpen
  \bibfield  {author} {\bibinfo {author} {\bibfnamefont {L.~F.}\ \bibnamefont
  {Alday}}\ and\ \bibinfo {author} {\bibfnamefont {X.}~\bibnamefont {Zhou}},\
  }\bibfield  {title} {\bibinfo {title} {{Simplicity of AdS Supergravity at One
  Loop}},\ }\href {https://doi.org/10.1007/JHEP09(2020)008} {\bibfield
  {journal} {\bibinfo  {journal} {JHEP}\ }\textbf {\bibinfo {volume} {09}},\
  \bibinfo {pages} {008}},\ \Eprint {https://arxiv.org/abs/1912.02663}
  {arXiv:1912.02663 [hep-th]} \BibitemShut {NoStop}%
\bibitem [{\citenamefont {Huang}\ and\ \citenamefont
  {Yuan}(2023)}]{Huang:2021xws}%
  \BibitemOpen
  \bibfield  {author} {\bibinfo {author} {\bibfnamefont {Z.}~\bibnamefont
  {Huang}}\ and\ \bibinfo {author} {\bibfnamefont {E.~Y.}\ \bibnamefont
  {Yuan}},\ }\bibfield  {title} {\bibinfo {title} {{Graviton scattering in
  AdS$_{5}$\texttimes{} S$^{5}$ at two loops}},\ }\href
  {https://doi.org/10.1007/JHEP04(2023)064} {\bibfield  {journal} {\bibinfo
  {journal} {JHEP}\ }\textbf {\bibinfo {volume} {04}},\ \bibinfo {pages}
  {064}},\ \Eprint {https://arxiv.org/abs/2112.15174} {arXiv:2112.15174
  [hep-th]} \BibitemShut {NoStop}%
\bibitem [{\citenamefont {Drummond}\ and\ \citenamefont
  {Paul}(2022)}]{Drummond:2022dxw}%
  \BibitemOpen
  \bibfield  {author} {\bibinfo {author} {\bibfnamefont {J.~M.}\ \bibnamefont
  {Drummond}}\ and\ \bibinfo {author} {\bibfnamefont {H.}~\bibnamefont
  {Paul}},\ }\bibfield  {title} {\bibinfo {title} {{Two-loop supergravity on
  AdS$_{5}$\texttimes{}S$^{5}$ from CFT}},\ }\href
  {https://doi.org/10.1007/JHEP08(2022)275} {\bibfield  {journal} {\bibinfo
  {journal} {JHEP}\ }\textbf {\bibinfo {volume} {08}},\ \bibinfo {pages}
  {275}},\ \Eprint {https://arxiv.org/abs/2204.01829} {arXiv:2204.01829
  [hep-th]} \BibitemShut {NoStop}%
\bibitem [{\citenamefont {Huang}\ \emph {et~al.}(2024)\citenamefont {Huang},
  \citenamefont {Wang},\ and\ \citenamefont {Yuan}}]{Huang:2024rxr}%
  \BibitemOpen
  \bibfield  {author} {\bibinfo {author} {\bibfnamefont {Z.}~\bibnamefont
  {Huang}}, \bibinfo {author} {\bibfnamefont {B.}~\bibnamefont {Wang}},\ and\
  \bibinfo {author} {\bibfnamefont {E.~Y.}\ \bibnamefont {Yuan}},\ }\bibfield
  {title} {\bibinfo {title} {{All Next-Next-to-Extremal One-Loop Correlators of
  AdS Supergluons and Supergravitons}},\ }\href@noop {} {\  (\bibinfo {year}
  {2024})},\ \Eprint {https://arxiv.org/abs/2407.03408} {arXiv:2407.03408
  [hep-th]} \BibitemShut {NoStop}%
\bibitem [{\citenamefont {Gon{\c{c}}alves}(2015)}]{Goncalves:2014ffa}%
  \BibitemOpen
  \bibfield  {author} {\bibinfo {author} {\bibfnamefont {V.}~\bibnamefont
  {Gon{\c{c}}alves}},\ }\bibfield  {title} {\bibinfo {title} {{Four point
  function of $\mathcal{N}=4$ stress-tensor multiplet at strong coupling}},\
  }\href {https://doi.org/10.1007/JHEP04(2015)150} {\bibfield  {journal}
  {\bibinfo  {journal} {JHEP}\ }\textbf {\bibinfo {volume} {04}},\ \bibinfo
  {pages} {150}},\ \Eprint {https://arxiv.org/abs/1411.1675} {arXiv:1411.1675
  [hep-th]} \BibitemShut {NoStop}%
\bibitem [{\citenamefont {Alday}\ \emph {et~al.}(2019)\citenamefont {Alday},
  \citenamefont {Bissi},\ and\ \citenamefont {Perlmutter}}]{Alday:2018pdi}%
  \BibitemOpen
  \bibfield  {author} {\bibinfo {author} {\bibfnamefont {L.~F.}\ \bibnamefont
  {Alday}}, \bibinfo {author} {\bibfnamefont {A.}~\bibnamefont {Bissi}},\ and\
  \bibinfo {author} {\bibfnamefont {E.}~\bibnamefont {Perlmutter}},\ }\bibfield
   {title} {\bibinfo {title} {{Genus-One String Amplitudes from Conformal Field
  Theory}},\ }\href {https://doi.org/10.1007/JHEP06(2019)010} {\bibfield
  {journal} {\bibinfo  {journal} {JHEP}\ }\textbf {\bibinfo {volume} {06}},\
  \bibinfo {pages} {010}},\ \Eprint {https://arxiv.org/abs/1809.10670}
  {arXiv:1809.10670 [hep-th]} \BibitemShut {NoStop}%
\bibitem [{\citenamefont {Alday}(2021)}]{Alday:2018kkw}%
  \BibitemOpen
  \bibfield  {author} {\bibinfo {author} {\bibfnamefont {L.~F.}\ \bibnamefont
  {Alday}},\ }\bibfield  {title} {\bibinfo {title} {{On genus-one string
  amplitudes on $AdS_5 \times S^5$}},\ }\href
  {https://doi.org/10.1007/JHEP04(2021)005} {\bibfield  {journal} {\bibinfo
  {journal} {JHEP}\ }\textbf {\bibinfo {volume} {04}},\ \bibinfo {pages}
  {005}},\ \Eprint {https://arxiv.org/abs/1812.11783} {arXiv:1812.11783
  [hep-th]} \BibitemShut {NoStop}%
\bibitem [{\citenamefont {Drummond}\ \emph {et~al.}(2019)\citenamefont
  {Drummond}, \citenamefont {Nandan}, \citenamefont {Paul},\ and\ \citenamefont
  {Rigatos}}]{Drummond:2019odu}%
  \BibitemOpen
  \bibfield  {author} {\bibinfo {author} {\bibfnamefont {J.~M.}\ \bibnamefont
  {Drummond}}, \bibinfo {author} {\bibfnamefont {D.}~\bibnamefont {Nandan}},
  \bibinfo {author} {\bibfnamefont {H.}~\bibnamefont {Paul}},\ and\ \bibinfo
  {author} {\bibfnamefont {K.~S.}\ \bibnamefont {Rigatos}},\ }\bibfield
  {title} {\bibinfo {title} {{String corrections to AdS amplitudes and the
  double-trace spectrum of $ \mathcal{N} $ = 4 SYM}},\ }\href
  {https://doi.org/10.1007/JHEP12(2019)173} {\bibfield  {journal} {\bibinfo
  {journal} {JHEP}\ }\textbf {\bibinfo {volume} {12}},\ \bibinfo {pages}
  {173}},\ \Eprint {https://arxiv.org/abs/1907.00992} {arXiv:1907.00992
  [hep-th]} \BibitemShut {NoStop}%
\bibitem [{\citenamefont {Drummond}\ and\ \citenamefont
  {Paul}(2021)}]{Drummond:2019hel}%
  \BibitemOpen
  \bibfield  {author} {\bibinfo {author} {\bibfnamefont {J.~M.}\ \bibnamefont
  {Drummond}}\ and\ \bibinfo {author} {\bibfnamefont {H.}~\bibnamefont
  {Paul}},\ }\bibfield  {title} {\bibinfo {title} {{One-loop string corrections
  to AdS amplitudes from CFT}},\ }\href
  {https://doi.org/10.1007/JHEP03(2021)038} {\bibfield  {journal} {\bibinfo
  {journal} {JHEP}\ }\textbf {\bibinfo {volume} {03}},\ \bibinfo {pages}
  {038}},\ \Eprint {https://arxiv.org/abs/1912.07632} {arXiv:1912.07632
  [hep-th]} \BibitemShut {NoStop}%
\bibitem [{\citenamefont {Drummond}\ \emph {et~al.}(2023)\citenamefont
  {Drummond}, \citenamefont {Paul},\ and\ \citenamefont
  {Santagata}}]{Drummond:2020dwr}%
  \BibitemOpen
  \bibfield  {author} {\bibinfo {author} {\bibfnamefont {J.~M.}\ \bibnamefont
  {Drummond}}, \bibinfo {author} {\bibfnamefont {H.}~\bibnamefont {Paul}},\
  and\ \bibinfo {author} {\bibfnamefont {M.}~\bibnamefont {Santagata}},\
  }\bibfield  {title} {\bibinfo {title} {{Bootstrapping string theory on
  AdS5{\texttimes}S5}},\ }\href {https://doi.org/10.1103/PhysRevD.108.026020}
  {\bibfield  {journal} {\bibinfo  {journal} {Phys. Rev. D}\ }\textbf {\bibinfo
  {volume} {108}},\ \bibinfo {pages} {026020} (\bibinfo {year} {2023})},\
  \Eprint {https://arxiv.org/abs/2004.07282} {arXiv:2004.07282 [hep-th]}
  \BibitemShut {NoStop}%
\bibitem [{\citenamefont {Drummond}\ \emph {et~al.}(2021)\citenamefont
  {Drummond}, \citenamefont {Glew},\ and\ \citenamefont
  {Paul}}]{Drummond:2020uni}%
  \BibitemOpen
  \bibfield  {author} {\bibinfo {author} {\bibfnamefont {J.~M.}\ \bibnamefont
  {Drummond}}, \bibinfo {author} {\bibfnamefont {R.}~\bibnamefont {Glew}},\
  and\ \bibinfo {author} {\bibfnamefont {H.}~\bibnamefont {Paul}},\ }\bibfield
  {title} {\bibinfo {title} {{One-loop string corrections for AdS Kaluza-Klein
  amplitudes}},\ }\href {https://doi.org/10.1007/JHEP12(2021)072} {\bibfield
  {journal} {\bibinfo  {journal} {JHEP}\ }\textbf {\bibinfo {volume} {12}},\
  \bibinfo {pages} {072}},\ \Eprint {https://arxiv.org/abs/2008.01109}
  {arXiv:2008.01109 [hep-th]} \BibitemShut {NoStop}%
\bibitem [{\citenamefont {Aprile}\ \emph {et~al.}(2021)\citenamefont {Aprile},
  \citenamefont {Drummond}, \citenamefont {Paul},\ and\ \citenamefont
  {Santagata}}]{Aprile:2020mus}%
  \BibitemOpen
  \bibfield  {author} {\bibinfo {author} {\bibfnamefont {F.}~\bibnamefont
  {Aprile}}, \bibinfo {author} {\bibfnamefont {J.~M.}\ \bibnamefont
  {Drummond}}, \bibinfo {author} {\bibfnamefont {H.}~\bibnamefont {Paul}},\
  and\ \bibinfo {author} {\bibfnamefont {M.}~\bibnamefont {Santagata}},\
  }\bibfield  {title} {\bibinfo {title} {{The Virasoro-Shapiro amplitude in
  AdS$_{5}$ \texttimes{} S$^{5}$ and level splitting of 10d conformal
  symmetry}},\ }\href {https://doi.org/10.1007/JHEP11(2021)109} {\bibfield
  {journal} {\bibinfo  {journal} {JHEP}\ }\textbf {\bibinfo {volume} {11}},\
  \bibinfo {pages} {109}},\ \Eprint {https://arxiv.org/abs/2012.12092}
  {arXiv:2012.12092 [hep-th]} \BibitemShut {NoStop}%
\bibitem [{\citenamefont {Aprile}\ \emph {et~al.}(2023)\citenamefont {Aprile},
  \citenamefont {Drummond}, \citenamefont {Glew},\ and\ \citenamefont
  {Santagata}}]{Aprile:2022tzr}%
  \BibitemOpen
  \bibfield  {author} {\bibinfo {author} {\bibfnamefont {F.}~\bibnamefont
  {Aprile}}, \bibinfo {author} {\bibfnamefont {J.~M.}\ \bibnamefont
  {Drummond}}, \bibinfo {author} {\bibfnamefont {R.}~\bibnamefont {Glew}},\
  and\ \bibinfo {author} {\bibfnamefont {M.}~\bibnamefont {Santagata}},\
  }\bibfield  {title} {\bibinfo {title} {{One-loop string amplitudes in
  AdS$_{5}$\texttimes{}S$^{5}$: Mellin space and sphere splitting}},\ }\href
  {https://doi.org/10.1007/JHEP02(2023)087} {\bibfield  {journal} {\bibinfo
  {journal} {JHEP}\ }\textbf {\bibinfo {volume} {02}},\ \bibinfo {pages}
  {087}},\ \Eprint {https://arxiv.org/abs/2207.13084} {arXiv:2207.13084
  [hep-th]} \BibitemShut {NoStop}%
\bibitem [{\citenamefont {Bissi}\ \emph {et~al.}(2022)\citenamefont {Bissi},
  \citenamefont {Sinha},\ and\ \citenamefont {Zhou}}]{Bissi:2022mrs}%
  \BibitemOpen
  \bibfield  {author} {\bibinfo {author} {\bibfnamefont {A.}~\bibnamefont
  {Bissi}}, \bibinfo {author} {\bibfnamefont {A.}~\bibnamefont {Sinha}},\ and\
  \bibinfo {author} {\bibfnamefont {X.}~\bibnamefont {Zhou}},\ }\bibfield
  {title} {\bibinfo {title} {{Selected topics in analytic conformal bootstrap:
  A guided journey}},\ }\href {https://doi.org/10.1016/j.physrep.2022.09.004}
  {\bibfield  {journal} {\bibinfo  {journal} {Phys. Rept.}\ }\textbf {\bibinfo
  {volume} {991}},\ \bibinfo {pages} {1} (\bibinfo {year} {2022})},\ \Eprint
  {https://arxiv.org/abs/2202.08475} {arXiv:2202.08475 [hep-th]} \BibitemShut
  {NoStop}%
\bibitem [{\citenamefont {Heslop}(2022)}]{Heslop:2022xgp}%
  \BibitemOpen
  \bibfield  {author} {\bibinfo {author} {\bibfnamefont {P.}~\bibnamefont
  {Heslop}},\ }\bibfield  {title} {\bibinfo {title} {{The SAGEX Review on
  Scattering Amplitudes, Chapter 8: Half BPS correlators}},\ }\href
  {https://doi.org/10.1088/1751-8121/ac8c71} {\bibfield  {journal} {\bibinfo
  {journal} {J. Phys. A}\ }\textbf {\bibinfo {volume} {55}},\ \bibinfo {pages}
  {443009} (\bibinfo {year} {2022})},\ \Eprint
  {https://arxiv.org/abs/2203.13019} {arXiv:2203.13019 [hep-th]} \BibitemShut
  {NoStop}%
\bibitem [{\citenamefont {Caron-Huot}\ and\ \citenamefont
  {Trinh}(2019)}]{Caron-Huot:2018kta}%
  \BibitemOpen
  \bibfield  {author} {\bibinfo {author} {\bibfnamefont {S.}~\bibnamefont
  {Caron-Huot}}\ and\ \bibinfo {author} {\bibfnamefont {A.-K.}\ \bibnamefont
  {Trinh}},\ }\bibfield  {title} {\bibinfo {title} {{All Tree-Level Correlators
  in AdS${}_5\times$S${}_5$ Supergravity: Hidden Ten-Dimensional Conformal
  Symmetry}},\ }\href {https://doi.org/10.1007/JHEP01(2019)196} {\bibfield
  {journal} {\bibinfo  {journal} {JHEP}\ }\textbf {\bibinfo {volume} {01}},\
  \bibinfo {pages} {196}},\ \Eprint {https://arxiv.org/abs/1809.09173}
  {arXiv:1809.09173 [hep-th]} \BibitemShut {NoStop}%
\bibitem [{\citenamefont {Abl}\ \emph {et~al.}(2021)\citenamefont {Abl},
  \citenamefont {Heslop},\ and\ \citenamefont {Lipstein}}]{Abl:2020dbx}%
  \BibitemOpen
  \bibfield  {author} {\bibinfo {author} {\bibfnamefont {T.}~\bibnamefont
  {Abl}}, \bibinfo {author} {\bibfnamefont {P.}~\bibnamefont {Heslop}},\ and\
  \bibinfo {author} {\bibfnamefont {A.~E.}\ \bibnamefont {Lipstein}},\
  }\bibfield  {title} {\bibinfo {title} {{Towards the Virasoro-Shapiro
  amplitude in AdS$_{5} \times S^{5}$}},\ }\href
  {https://doi.org/10.1007/JHEP04(2021)237} {\bibfield  {journal} {\bibinfo
  {journal} {JHEP}\ }\textbf {\bibinfo {volume} {04}},\ \bibinfo {pages}
  {237}},\ \Eprint {https://arxiv.org/abs/2012.12091} {arXiv:2012.12091
  [hep-th]} \BibitemShut {NoStop}%
\bibitem [{\citenamefont {Caron-Huot}\ and\ \citenamefont
  {Coronado}(2022)}]{Caron-Huot:2021usw}%
  \BibitemOpen
  \bibfield  {author} {\bibinfo {author} {\bibfnamefont {S.}~\bibnamefont
  {Caron-Huot}}\ and\ \bibinfo {author} {\bibfnamefont {F.}~\bibnamefont
  {Coronado}},\ }\bibfield  {title} {\bibinfo {title} {{Ten dimensional
  symmetry of $ \mathcal{N} $ = 4 SYM correlators}},\ }\href
  {https://doi.org/10.1007/JHEP03(2022)151} {\bibfield  {journal} {\bibinfo
  {journal} {JHEP}\ }\textbf {\bibinfo {volume} {03}},\ \bibinfo {pages}
  {151}},\ \Eprint {https://arxiv.org/abs/2106.03892} {arXiv:2106.03892
  [hep-th]} \BibitemShut {NoStop}%
\bibitem [{\citenamefont {Caron-Huot}\ \emph {et~al.}(2023)\citenamefont
  {Caron-Huot}, \citenamefont {Coronado},\ and\ \citenamefont
  {M{\"u}hlmann}}]{Caron-Huot:2023wdh}%
  \BibitemOpen
  \bibfield  {author} {\bibinfo {author} {\bibfnamefont {S.}~\bibnamefont
  {Caron-Huot}}, \bibinfo {author} {\bibfnamefont {F.}~\bibnamefont
  {Coronado}},\ and\ \bibinfo {author} {\bibfnamefont {B.}~\bibnamefont
  {M{\"u}hlmann}},\ }\bibfield  {title} {\bibinfo {title} {{Determinants in
  self-dual $ \mathcal{N} $ = 4 SYM and twistor space}},\ }\href
  {https://doi.org/10.1007/JHEP08(2023)008} {\bibfield  {journal} {\bibinfo
  {journal} {JHEP}\ }\textbf {\bibinfo {volume} {08}},\ \bibinfo {pages}
  {008}},\ \Eprint {https://arxiv.org/abs/2304.12341} {arXiv:2304.12341
  [hep-th]} \BibitemShut {NoStop}%
\bibitem [{\citenamefont {Aprile}\ and\ \citenamefont
  {Vieira}(2020)}]{Aprile:2020luw}%
  \BibitemOpen
  \bibfield  {author} {\bibinfo {author} {\bibfnamefont {F.}~\bibnamefont
  {Aprile}}\ and\ \bibinfo {author} {\bibfnamefont {P.}~\bibnamefont
  {Vieira}},\ }\bibfield  {title} {\bibinfo {title} {{Large $p$ explorations.
  From SUGRA to big STRINGS in Mellin space}},\ }\href
  {https://doi.org/10.1007/JHEP12(2020)206} {\bibfield  {journal} {\bibinfo
  {journal} {JHEP}\ }\textbf {\bibinfo {volume} {12}},\ \bibinfo {pages}
  {206}},\ \Eprint {https://arxiv.org/abs/2007.09176} {arXiv:2007.09176
  [hep-th]} \BibitemShut {NoStop}%
\bibitem [{\citenamefont {Huang}\ \emph {et~al.}(2025)\citenamefont {Huang},
  \citenamefont {Wang}, \citenamefont {Yuan},\ and\ \citenamefont
  {Zhang}}]{Huang:2024dxr}%
  \BibitemOpen
  \bibfield  {author} {\bibinfo {author} {\bibfnamefont {Z.}~\bibnamefont
  {Huang}}, \bibinfo {author} {\bibfnamefont {B.}~\bibnamefont {Wang}},
  \bibinfo {author} {\bibfnamefont {E.~Y.}\ \bibnamefont {Yuan}},\ and\
  \bibinfo {author} {\bibfnamefont {J.}~\bibnamefont {Zhang}},\ }\bibfield
  {title} {\bibinfo {title} {{All Five-Point Kaluza-Klein Correlators and
  Hidden 8D Symmetry in AdS5{\texttimes}S3}},\ }\href
  {https://doi.org/10.1103/PhysRevLett.134.161601} {\bibfield  {journal}
  {\bibinfo  {journal} {Phys. Rev. Lett.}\ }\textbf {\bibinfo {volume} {134}},\
  \bibinfo {pages} {161601} (\bibinfo {year} {2025})},\ \Eprint
  {https://arxiv.org/abs/2408.12260} {arXiv:2408.12260 [hep-th]} \BibitemShut
  {NoStop}%
\bibitem [{\citenamefont {Wang}\ \emph {et~al.}(2025)\citenamefont {Wang},
  \citenamefont {Wu},\ and\ \citenamefont {Yuan}}]{Wang:2025pjo}%
  \BibitemOpen
  \bibfield  {author} {\bibinfo {author} {\bibfnamefont {B.}~\bibnamefont
  {Wang}}, \bibinfo {author} {\bibfnamefont {D.}~\bibnamefont {Wu}},\ and\
  \bibinfo {author} {\bibfnamefont {E.~Y.}\ \bibnamefont {Yuan}},\ }\bibfield
  {title} {\bibinfo {title} {{The Kaluza-Klein AdS Virasoro-Shapiro Amplitude
  near Flat Space}},\ }\href@noop {} {\  (\bibinfo {year} {2025})},\ \Eprint
  {https://arxiv.org/abs/2503.01964} {arXiv:2503.01964 [hep-th]} \BibitemShut
  {NoStop}%
\bibitem [{Note1()}]{Note1}%
  \BibitemOpen
  \bibinfo {note} {See also \cite {Chen:2025yxg,Wu:2025ott} for recent studies
  on hidden 10d symmetry in giant graviton correlators.}\BibitemShut {Stop}%
\bibitem [{\citenamefont {Alday}\ \emph {et~al.}(2022)\citenamefont {Alday},
  \citenamefont {Gon\c{c}alves},\ and\ \citenamefont {Zhou}}]{Alday:2022lkk}%
  \BibitemOpen
  \bibfield  {author} {\bibinfo {author} {\bibfnamefont {L.~F.}\ \bibnamefont
  {Alday}}, \bibinfo {author} {\bibfnamefont {V.}~\bibnamefont
  {Gon\c{c}alves}},\ and\ \bibinfo {author} {\bibfnamefont {X.}~\bibnamefont
  {Zhou}},\ }\bibfield  {title} {\bibinfo {title} {{Supersymmetric Five-Point
  Gluon Amplitudes in AdS Space}},\ }\href
  {https://doi.org/10.1103/PhysRevLett.128.161601} {\bibfield  {journal}
  {\bibinfo  {journal} {Phys. Rev. Lett.}\ }\textbf {\bibinfo {volume} {128}},\
  \bibinfo {pages} {161601} (\bibinfo {year} {2022})},\ \Eprint
  {https://arxiv.org/abs/2201.04422} {arXiv:2201.04422 [hep-th]} \BibitemShut
  {NoStop}%
\bibitem [{\citenamefont {Alday}\ \emph {et~al.}(2024)\citenamefont {Alday},
  \citenamefont {Gon{\c{c}}alves}, \citenamefont {Nocchi},\ and\ \citenamefont
  {Zhou}}]{Alday:2023kfm}%
  \BibitemOpen
  \bibfield  {author} {\bibinfo {author} {\bibfnamefont {L.~F.}\ \bibnamefont
  {Alday}}, \bibinfo {author} {\bibfnamefont {V.}~\bibnamefont
  {Gon{\c{c}}alves}}, \bibinfo {author} {\bibfnamefont {M.}~\bibnamefont
  {Nocchi}},\ and\ \bibinfo {author} {\bibfnamefont {X.}~\bibnamefont {Zhou}},\
  }\bibfield  {title} {\bibinfo {title} {{Six-point AdS gluon amplitudes from
  flat space and factorization}},\ }\href
  {https://doi.org/10.1103/PhysRevResearch.6.L012041} {\bibfield  {journal}
  {\bibinfo  {journal} {Phys. Rev. Res.}\ }\textbf {\bibinfo {volume} {6}},\
  \bibinfo {pages} {L012041} (\bibinfo {year} {2024})},\ \Eprint
  {https://arxiv.org/abs/2307.06884} {arXiv:2307.06884 [hep-th]} \BibitemShut
  {NoStop}%
\bibitem [{\citenamefont {Cao}\ \emph {et~al.}(2024{\natexlab{a}})\citenamefont
  {Cao}, \citenamefont {He},\ and\ \citenamefont {Tang}}]{Cao:2023cwa}%
  \BibitemOpen
  \bibfield  {author} {\bibinfo {author} {\bibfnamefont {Q.}~\bibnamefont
  {Cao}}, \bibinfo {author} {\bibfnamefont {S.}~\bibnamefont {He}},\ and\
  \bibinfo {author} {\bibfnamefont {Y.}~\bibnamefont {Tang}},\ }\bibfield
  {title} {\bibinfo {title} {{Constructibility of AdS Supergluon Amplitudes}},\
  }\href {https://doi.org/10.1103/PhysRevLett.133.021605} {\bibfield  {journal}
  {\bibinfo  {journal} {Phys. Rev. Lett.}\ }\textbf {\bibinfo {volume} {133}},\
  \bibinfo {pages} {021605} (\bibinfo {year} {2024}{\natexlab{a}})},\ \Eprint
  {https://arxiv.org/abs/2312.15484} {arXiv:2312.15484 [hep-th]} \BibitemShut
  {NoStop}%
\bibitem [{\citenamefont {Cao}\ \emph {et~al.}(2024{\natexlab{b}})\citenamefont
  {Cao}, \citenamefont {He}, \citenamefont {Li},\ and\ \citenamefont
  {Tang}}]{Cao:2024bky}%
  \BibitemOpen
  \bibfield  {author} {\bibinfo {author} {\bibfnamefont {Q.}~\bibnamefont
  {Cao}}, \bibinfo {author} {\bibfnamefont {S.}~\bibnamefont {He}}, \bibinfo
  {author} {\bibfnamefont {X.}~\bibnamefont {Li}},\ and\ \bibinfo {author}
  {\bibfnamefont {Y.}~\bibnamefont {Tang}},\ }\bibfield  {title} {\bibinfo
  {title} {{Supergluon scattering in AdS: constructibility, spinning
  amplitudes, and new structures}},\ }\href@noop {} {\  (\bibinfo {year}
  {2024}{\natexlab{b}})},\ \Eprint {https://arxiv.org/abs/2406.08538}
  {arXiv:2406.08538 [hep-th]} \BibitemShut {NoStop}%
\bibitem [{\citenamefont {Gon{\c c}alves}\ \emph {et~al.}(2019)\citenamefont
  {Gon{\c c}alves}, \citenamefont {Pereira},\ and\ \citenamefont
  {Zhou}}]{Goncalves:2019znr}%
  \BibitemOpen
  \bibfield  {author} {\bibinfo {author} {\bibfnamefont {V.}~\bibnamefont
  {Gon{\c c}alves}}, \bibinfo {author} {\bibfnamefont {R.}~\bibnamefont
  {Pereira}},\ and\ \bibinfo {author} {\bibfnamefont {X.}~\bibnamefont
  {Zhou}},\ }\bibfield  {title} {\bibinfo {title} {{$20'$ Five-Point Function
  from $AdS_5\times S^5$ Supergravity}},\ }\href
  {https://doi.org/10.1007/JHEP10(2019)247} {\bibfield  {journal} {\bibinfo
  {journal} {JHEP}\ }\textbf {\bibinfo {volume} {10}},\ \bibinfo {pages}
  {247}},\ \Eprint {https://arxiv.org/abs/1906.05305} {arXiv:1906.05305
  [hep-th]} \BibitemShut {NoStop}%
\bibitem [{\citenamefont {Gon\c{c}alves}\ \emph {et~al.}(2023)\citenamefont
  {Gon\c{c}alves}, \citenamefont {Meneghelli}, \citenamefont {Pereira},
  \citenamefont {Vilas~Boas},\ and\ \citenamefont {Zhou}}]{Goncalves:2023oyx}%
  \BibitemOpen
  \bibfield  {author} {\bibinfo {author} {\bibfnamefont {V.}~\bibnamefont
  {Gon\c{c}alves}}, \bibinfo {author} {\bibfnamefont {C.}~\bibnamefont
  {Meneghelli}}, \bibinfo {author} {\bibfnamefont {R.}~\bibnamefont {Pereira}},
  \bibinfo {author} {\bibfnamefont {J.}~\bibnamefont {Vilas~Boas}},\ and\
  \bibinfo {author} {\bibfnamefont {X.}~\bibnamefont {Zhou}},\ }\bibfield
  {title} {\bibinfo {title} {{Kaluza-Klein five-point functions from
  AdS$_{5}$\texttimes{}S$^{5}$ supergravity}},\ }\href
  {https://doi.org/10.1007/JHEP08(2023)067} {\bibfield  {journal} {\bibinfo
  {journal} {JHEP}\ }\textbf {\bibinfo {volume} {08}},\ \bibinfo {pages}
  {067}},\ \Eprint {https://arxiv.org/abs/2302.01896} {arXiv:2302.01896
  [hep-th]} \BibitemShut {NoStop}%
\bibitem [{\citenamefont {Goncalves}\ \emph {et~al.}(2025)\citenamefont
  {Goncalves}, \citenamefont {Nocchi},\ and\ \citenamefont
  {Zhou}}]{Goncalves:2025jcg}%
  \BibitemOpen
  \bibfield  {author} {\bibinfo {author} {\bibfnamefont {V.}~\bibnamefont
  {Goncalves}}, \bibinfo {author} {\bibfnamefont {M.}~\bibnamefont {Nocchi}},\
  and\ \bibinfo {author} {\bibfnamefont {X.}~\bibnamefont {Zhou}},\ }\bibfield
  {title} {\bibinfo {title} {{Dissecting supergraviton six-point function with
  lightcone limits and chiral algebra}},\ }\href
  {https://doi.org/10.1007/JHEP06(2025)173} {\bibfield  {journal} {\bibinfo
  {journal} {JHEP}\ }\textbf {\bibinfo {volume} {06}},\ \bibinfo {pages}
  {173}},\ \Eprint {https://arxiv.org/abs/2502.10269} {arXiv:2502.10269
  [hep-th]} \BibitemShut {NoStop}%
\bibitem [{\citenamefont {Drukker}\ and\ \citenamefont
  {Plefka}(2009)}]{Drukker:2009sf}%
  \BibitemOpen
  \bibfield  {author} {\bibinfo {author} {\bibfnamefont {N.}~\bibnamefont
  {Drukker}}\ and\ \bibinfo {author} {\bibfnamefont {J.}~\bibnamefont
  {Plefka}},\ }\bibfield  {title} {\bibinfo {title} {{Superprotected n-point
  correlation functions of local operators in N=4 super Yang-Mills}},\ }\href
  {https://doi.org/10.1088/1126-6708/2009/04/052} {\bibfield  {journal}
  {\bibinfo  {journal} {JHEP}\ }\textbf {\bibinfo {volume} {04}},\ \bibinfo
  {pages} {052}},\ \Eprint {https://arxiv.org/abs/0901.3653} {arXiv:0901.3653
  [hep-th]} \BibitemShut {NoStop}%
\bibitem [{\citenamefont {Costa}\ \emph {et~al.}(2011)\citenamefont {Costa},
  \citenamefont {Penedones}, \citenamefont {Poland},\ and\ \citenamefont
  {Rychkov}}]{Costa:2011mg}%
  \BibitemOpen
  \bibfield  {author} {\bibinfo {author} {\bibfnamefont {M.~S.}\ \bibnamefont
  {Costa}}, \bibinfo {author} {\bibfnamefont {J.}~\bibnamefont {Penedones}},
  \bibinfo {author} {\bibfnamefont {D.}~\bibnamefont {Poland}},\ and\ \bibinfo
  {author} {\bibfnamefont {S.}~\bibnamefont {Rychkov}},\ }\bibfield  {title}
  {\bibinfo {title} {{Spinning Conformal Correlators}},\ }\href
  {https://doi.org/10.1007/JHEP11(2011)071} {\bibfield  {journal} {\bibinfo
  {journal} {JHEP}\ }\textbf {\bibinfo {volume} {11}},\ \bibinfo {pages}
  {071}},\ \Eprint {https://arxiv.org/abs/1107.3554} {arXiv:1107.3554 [hep-th]}
  \BibitemShut {NoStop}%
\bibitem [{Note2()}]{Note2}%
  \BibitemOpen
  \bibinfo {note} {For simplicity, in the main text we present only the
  discussion of scalar correlators. The spinning cases can be found in the
  Supplemental Material.}\BibitemShut {Stop}%
\bibitem [{Note3()}]{Note3}%
  \BibitemOpen
  \bibinfo {note} {Due to the $1/n_{ij}!$ factors, the sum is automatically
  truncated to a finite number of integer points such that all $n_{ij}\geq
  0$.}\BibitemShut {Stop}%
\bibitem [{\citenamefont {Dolan}\ \emph {et~al.}(2004)\citenamefont {Dolan},
  \citenamefont {Gallot},\ and\ \citenamefont {Sokatchev}}]{Dolan:2004mu}%
  \BibitemOpen
  \bibfield  {author} {\bibinfo {author} {\bibfnamefont {F.~A.}\ \bibnamefont
  {Dolan}}, \bibinfo {author} {\bibfnamefont {L.}~\bibnamefont {Gallot}},\ and\
  \bibinfo {author} {\bibfnamefont {E.}~\bibnamefont {Sokatchev}},\ }\bibfield
  {title} {\bibinfo {title} {{On four-point functions of 1/2-BPS operators in
  general dimensions}},\ }\href {https://doi.org/10.1088/1126-6708/2004/09/056}
  {\bibfield  {journal} {\bibinfo  {journal} {JHEP}\ }\textbf {\bibinfo
  {volume} {09}},\ \bibinfo {pages} {056}},\ \Eprint
  {https://arxiv.org/abs/hep-th/0405180} {arXiv:hep-th/0405180} \BibitemShut
  {NoStop}%
\bibitem [{\citenamefont {Gon{\c c}alves}\ \emph {et~al.}(2015)\citenamefont
  {Gon{\c c}alves}, \citenamefont {Penedones},\ and\ \citenamefont
  {Trevisani}}]{Goncalves:2014rfa}%
  \BibitemOpen
  \bibfield  {author} {\bibinfo {author} {\bibfnamefont {V.}~\bibnamefont
  {Gon{\c c}alves}}, \bibinfo {author} {\bibfnamefont {J.}~\bibnamefont
  {Penedones}},\ and\ \bibinfo {author} {\bibfnamefont {E.}~\bibnamefont
  {Trevisani}},\ }\bibfield  {title} {\bibinfo {title} {{Factorization of
  Mellin amplitudes}},\ }\href {https://doi.org/10.1007/JHEP10(2015)040}
  {\bibfield  {journal} {\bibinfo  {journal} {JHEP}\ }\textbf {\bibinfo
  {volume} {10}},\ \bibinfo {pages} {040}},\ \Eprint
  {https://arxiv.org/abs/1410.4185} {arXiv:1410.4185 [hep-th]} \BibitemShut
  {NoStop}%
\bibitem [{Note4()}]{Note4}%
  \BibitemOpen
  \bibinfo {note} {At tree level, multi-trace exchanges are encoded in poles of
  $\Gamma $ functions in the conventional Mellin transformation.}\BibitemShut
  {Stop}%
\bibitem [{\citenamefont {Fitzpatrick}\ \emph {et~al.}(2011)\citenamefont
  {Fitzpatrick}, \citenamefont {Kaplan}, \citenamefont {Penedones},
  \citenamefont {Raju},\ and\ \citenamefont {van Rees}}]{Fitzpatrick:2011ia}%
  \BibitemOpen
  \bibfield  {author} {\bibinfo {author} {\bibfnamefont {A.~L.}\ \bibnamefont
  {Fitzpatrick}}, \bibinfo {author} {\bibfnamefont {J.}~\bibnamefont {Kaplan}},
  \bibinfo {author} {\bibfnamefont {J.}~\bibnamefont {Penedones}}, \bibinfo
  {author} {\bibfnamefont {S.}~\bibnamefont {Raju}},\ and\ \bibinfo {author}
  {\bibfnamefont {B.~C.}\ \bibnamefont {van Rees}},\ }\bibfield  {title}
  {\bibinfo {title} {{A Natural Language for AdS/CFT Correlators}},\ }\href
  {https://doi.org/10.1007/JHEP11(2011)095} {\bibfield  {journal} {\bibinfo
  {journal} {JHEP}\ }\textbf {\bibinfo {volume} {11}},\ \bibinfo {pages}
  {095}},\ \Eprint {https://arxiv.org/abs/1107.1499} {arXiv:1107.1499 [hep-th]}
  \BibitemShut {NoStop}%
\bibitem [{Note5()}]{Note5}%
  \BibitemOpen
  \bibinfo {note} {When limited on the reduced correlators, this $Z_2$ symmetry
  can be viewed as an element in the hidden 10d conformal group.}\BibitemShut
  {Stop}%
\bibitem [{Note6()}]{Note6}%
  \BibitemOpen
  \bibinfo {note} {The same $Z_2$ symmetry was discovered in \cite
  {Aprile:2020mus} for stringy corrections. We thank Francesco Aprile and
  Michele Santagata for pointing out this.}\BibitemShut {Stop}%
\bibitem [{Note7()}]{Note7}%
  \BibitemOpen
  \bibinfo {note} {The amplitude $\protect \mathcal {M}_{3,\protect \mathcal
  {O}}$ might appear to violate the $Z_2$ symmetry $\gamma _{ij}\leftrightarrow
  -n_{ij}$, but in this three-point amplitude we have $\gamma _{ij}\equiv
  n_{ij}$.}\BibitemShut {Stop}%
\bibitem [{Note8()}]{Note8}%
  \BibitemOpen
  \bibinfo {note} {We will see that, after the bootstrap computation, the
  result can indeed be put in a factorized form $\mathcal {M}=\sum \mhat
  {R}\circ \widetilde {\mathcal M}$ with superconformal invariants $R$ and
  reduced amplitudes $\widetilde {\mathcal {M}}$ depending only on $\rho
  $.}\BibitemShut {Stop}%
\bibitem [{Note9()}]{Note9}%
  \BibitemOpen
  \bibinfo {note} {In fact, we did not write down an ansatz for $P^{(k_1,k_2)}$
  and $P^{(k_3)}$ explicitly in the actual bootstrap computation. All
  constraints on these polynomials should be understood as imposed directly on
  the RHS of \protect \eqref {eq:5ptfac1} and \protect \eqref
  {eq:5ptfac2}.}\BibitemShut {Stop}%
\bibitem [{Note10()}]{Note10}%
  \BibitemOpen
  \bibinfo {note} {See \cite {Heslop:2022xgp} and Supplemental Material for the
  way to construct these superconformal invariants.}\BibitemShut {Stop}%
\bibitem [{\citenamefont {Ceplak}\ \emph {et~al.}(2021)\citenamefont {Ceplak},
  \citenamefont {Giusto}, \citenamefont {Hughes},\ and\ \citenamefont
  {Russo}}]{Ceplak:2021wzz}%
  \BibitemOpen
  \bibfield  {author} {\bibinfo {author} {\bibfnamefont {N.}~\bibnamefont
  {Ceplak}}, \bibinfo {author} {\bibfnamefont {S.}~\bibnamefont {Giusto}},
  \bibinfo {author} {\bibfnamefont {M.~R.~R.}\ \bibnamefont {Hughes}},\ and\
  \bibinfo {author} {\bibfnamefont {R.}~\bibnamefont {Russo}},\ }\bibfield
  {title} {\bibinfo {title} {{Holographic correlators with multi-particle
  states}},\ }\href {https://doi.org/10.1007/JHEP09(2021)204} {\bibfield
  {journal} {\bibinfo  {journal} {JHEP}\ }\textbf {\bibinfo {volume} {09}},\
  \bibinfo {pages} {204}},\ \Eprint {https://arxiv.org/abs/2105.04670}
  {arXiv:2105.04670 [hep-th]} \BibitemShut {NoStop}%
\bibitem [{\citenamefont {Ma}\ and\ \citenamefont {Zhou}(2022)}]{Ma:2022ihn}%
  \BibitemOpen
  \bibfield  {author} {\bibinfo {author} {\bibfnamefont {W.-J.}\ \bibnamefont
  {Ma}}\ and\ \bibinfo {author} {\bibfnamefont {X.}~\bibnamefont {Zhou}},\
  }\bibfield  {title} {\bibinfo {title} {{Scattering bound states in AdS}},\
  }\href {https://doi.org/10.1007/JHEP08(2022)107} {\bibfield  {journal}
  {\bibinfo  {journal} {JHEP}\ }\textbf {\bibinfo {volume} {08}},\ \bibinfo
  {pages} {107}},\ \Eprint {https://arxiv.org/abs/2204.13419} {arXiv:2204.13419
  [hep-th]} \BibitemShut {NoStop}%
\bibitem [{\citenamefont {Fardelli}\ \emph {et~al.}(2024)\citenamefont
  {Fardelli}, \citenamefont {Fitzpatrick},\ and\ \citenamefont
  {Li}}]{Fardelli:2024heb}%
  \BibitemOpen
  \bibfield  {author} {\bibinfo {author} {\bibfnamefont {G.}~\bibnamefont
  {Fardelli}}, \bibinfo {author} {\bibfnamefont {A.~L.}\ \bibnamefont
  {Fitzpatrick}},\ and\ \bibinfo {author} {\bibfnamefont {W.}~\bibnamefont
  {Li}},\ }\bibfield  {title} {\bibinfo {title} {{Holography and Regge phases
  with U(1) charge}},\ }\href {https://doi.org/10.1007/JHEP08(2024)202}
  {\bibfield  {journal} {\bibinfo  {journal} {JHEP}\ }\textbf {\bibinfo
  {volume} {08}},\ \bibinfo {pages} {202}},\ \Eprint
  {https://arxiv.org/abs/2403.07079} {arXiv:2403.07079 [hep-th]} \BibitemShut
  {NoStop}%
\bibitem [{\citenamefont {Aprile}\ \emph
  {et~al.}(2025{\natexlab{a}})\citenamefont {Aprile}, \citenamefont {Giusto},\
  and\ \citenamefont {Russo}}]{Aprile:2024lwy}%
  \BibitemOpen
  \bibfield  {author} {\bibinfo {author} {\bibfnamefont {F.}~\bibnamefont
  {Aprile}}, \bibinfo {author} {\bibfnamefont {S.}~\bibnamefont {Giusto}},\
  and\ \bibinfo {author} {\bibfnamefont {R.}~\bibnamefont {Russo}},\ }\bibfield
   {title} {\bibinfo {title} {{Holographic correlators with BPS bound states in
  $\mathcal{N} = 4$ SYM}},\ }\href
  {https://doi.org/10.1103/PhysRevLett.134.091602} {\bibfield  {journal}
  {\bibinfo  {journal} {Phys. Rev. Lett.}\ }\textbf {\bibinfo {volume} {134}},\
  \bibinfo {pages} {091602} (\bibinfo {year} {2025}{\natexlab{a}})},\ \Eprint
  {https://arxiv.org/abs/2409.12911} {arXiv:2409.12911 [hep-th]} \BibitemShut
  {NoStop}%
\bibitem [{\citenamefont {Kravchuk}\ and\ \citenamefont
  {Mann}(2024)}]{Kravchuk:2024wmv}%
  \BibitemOpen
  \bibfield  {author} {\bibinfo {author} {\bibfnamefont {P.}~\bibnamefont
  {Kravchuk}}\ and\ \bibinfo {author} {\bibfnamefont {J.~A.}\ \bibnamefont
  {Mann}},\ }\bibfield  {title} {\bibinfo {title} {{AdS $N$-body problem at
  large spin}},\ }\href@noop {} {\  (\bibinfo {year} {2024})},\ \Eprint
  {https://arxiv.org/abs/2412.12328} {arXiv:2412.12328 [hep-th]} \BibitemShut
  {NoStop}%
\bibitem [{\citenamefont {Bissi}\ \emph {et~al.}(2025)\citenamefont {Bissi},
  \citenamefont {Fardelli},\ and\ \citenamefont {Manenti}}]{Bissi:2024tqf}%
  \BibitemOpen
  \bibfield  {author} {\bibinfo {author} {\bibfnamefont {A.}~\bibnamefont
  {Bissi}}, \bibinfo {author} {\bibfnamefont {G.}~\bibnamefont {Fardelli}},\
  and\ \bibinfo {author} {\bibfnamefont {A.}~\bibnamefont {Manenti}},\
  }\bibfield  {title} {\bibinfo {title} {{Composite operators in $ \mathcal{N}
  $ = 4 Super Yang-Mills}},\ }\href {https://doi.org/10.1007/JHEP07(2025)074}
  {\bibfield  {journal} {\bibinfo  {journal} {JHEP}\ }\textbf {\bibinfo
  {volume} {07}},\ \bibinfo {pages} {074}},\ \Eprint
  {https://arxiv.org/abs/2412.19788} {arXiv:2412.19788 [hep-th]} \BibitemShut
  {NoStop}%
\bibitem [{\citenamefont {Aprile}\ \emph
  {et~al.}(2025{\natexlab{b}})\citenamefont {Aprile}, \citenamefont {Giusto},\
  and\ \citenamefont {Russo}}]{Aprile:2025hlt}%
  \BibitemOpen
  \bibfield  {author} {\bibinfo {author} {\bibfnamefont {F.}~\bibnamefont
  {Aprile}}, \bibinfo {author} {\bibfnamefont {S.}~\bibnamefont {Giusto}},\
  and\ \bibinfo {author} {\bibfnamefont {R.}~\bibnamefont {Russo}},\ }\bibfield
   {title} {\bibinfo {title} {{Four-point correlators with BPS bound states in
  AdS$_3$ and AdS$_5$}},\ }\href@noop {} {\  (\bibinfo {year}
  {2025}{\natexlab{b}})},\ \Eprint {https://arxiv.org/abs/2503.02855}
  {arXiv:2503.02855 [hep-th]} \BibitemShut {NoStop}%
\bibitem [{\citenamefont {Vilas~Boas}(2025)}]{VilasBoas:2025vvw}%
  \BibitemOpen
  \bibfield  {author} {\bibinfo {author} {\bibfnamefont {J.}~\bibnamefont
  {Vilas~Boas}},\ }\bibfield  {title} {\bibinfo {title} {{$20'$ Five-Point
  Function of $\mathcal{N}=4$ SYM and Stringy Corrections}},\ }\href@noop {} {\
   (\bibinfo {year} {2025})},\ \Eprint {https://arxiv.org/abs/2507.12533}
  {arXiv:2507.12533 [hep-th]} \BibitemShut {NoStop}%
\bibitem [{\citenamefont {Bargheer}\ \emph {et~al.}(2023)\citenamefont
  {Bargheer}, \citenamefont {Fleury},\ and\ \citenamefont
  {Gon{\c{c}}alves}}]{Bargheer:2022sfd}%
  \BibitemOpen
  \bibfield  {author} {\bibinfo {author} {\bibfnamefont {T.}~\bibnamefont
  {Bargheer}}, \bibinfo {author} {\bibfnamefont {T.}~\bibnamefont {Fleury}},\
  and\ \bibinfo {author} {\bibfnamefont {V.}~\bibnamefont {Gon{\c{c}}alves}},\
  }\bibfield  {title} {\bibinfo {title} {{Higher-point integrands in
  $\mathcal{N} = 4$ super Yang-Mills theory}},\ }\href
  {https://doi.org/10.21468/SciPostPhys.15.2.059} {\bibfield  {journal}
  {\bibinfo  {journal} {SciPost Phys.}\ }\textbf {\bibinfo {volume} {15}},\
  \bibinfo {pages} {059} (\bibinfo {year} {2023})},\ \Eprint
  {https://arxiv.org/abs/2212.03773} {arXiv:2212.03773 [hep-th]} \BibitemShut
  {NoStop}%
\bibitem [{\citenamefont {Bargheer}\ \emph {et~al.}(2025)\citenamefont
  {Bargheer}, \citenamefont {Bekov}, \citenamefont {Bercini},\ and\
  \citenamefont {Coronado}}]{Bargheer:2025uai}%
  \BibitemOpen
  \bibfield  {author} {\bibinfo {author} {\bibfnamefont {T.}~\bibnamefont
  {Bargheer}}, \bibinfo {author} {\bibfnamefont {A.}~\bibnamefont {Bekov}},
  \bibinfo {author} {\bibfnamefont {C.}~\bibnamefont {Bercini}},\ and\ \bibinfo
  {author} {\bibfnamefont {F.}~\bibnamefont {Coronado}},\ }\bibfield  {title}
  {\bibinfo {title} {{Higher-Point Correlators in N=4 SYM: Generating
  Functions}},\ }\href@noop {} {\  (\bibinfo {year} {2025})},\ \Eprint
  {https://arxiv.org/abs/2509.14332} {arXiv:2509.14332 [hep-th]} \BibitemShut
  {NoStop}%
\bibitem [{\citenamefont {Zhou}(2021)}]{Zhou:2021gnu}%
  \BibitemOpen
  \bibfield  {author} {\bibinfo {author} {\bibfnamefont {X.}~\bibnamefont
  {Zhou}},\ }\bibfield  {title} {\bibinfo {title} {{Double Copy Relation in AdS
  Space}},\ }\href {https://doi.org/10.1103/PhysRevLett.127.141601} {\bibfield
  {journal} {\bibinfo  {journal} {Phys. Rev. Lett.}\ }\textbf {\bibinfo
  {volume} {127}},\ \bibinfo {pages} {141601} (\bibinfo {year} {2021})},\
  \Eprint {https://arxiv.org/abs/2106.07651} {arXiv:2106.07651 [hep-th]}
  \BibitemShut {NoStop}%
\bibitem [{\citenamefont {Chen}\ \emph {et~al.}(2025)\citenamefont {Chen},
  \citenamefont {Jiang},\ and\ \citenamefont {Zhou}}]{Chen:2025yxg}%
  \BibitemOpen
  \bibfield  {author} {\bibinfo {author} {\bibfnamefont {J.}~\bibnamefont
  {Chen}}, \bibinfo {author} {\bibfnamefont {Y.}~\bibnamefont {Jiang}},\ and\
  \bibinfo {author} {\bibfnamefont {X.}~\bibnamefont {Zhou}},\ }\bibfield
  {title} {\bibinfo {title} {{Giant Graviton Correlators as Defect Systems}},\
  }\href {https://doi.org/10.1103/hg9p-hblr} {\bibfield  {journal} {\bibinfo
  {journal} {Phys. Rev. Lett.}\ }\textbf {\bibinfo {volume} {135}},\ \bibinfo
  {pages} {081602} (\bibinfo {year} {2025})},\ \Eprint
  {https://arxiv.org/abs/2503.22987} {arXiv:2503.22987 [hep-th]} \BibitemShut
  {NoStop}%
\bibitem [{\citenamefont {Wu}\ \emph {et~al.}(2025)\citenamefont {Wu},
  \citenamefont {Jiang}, \citenamefont {Liu},\ and\ \citenamefont
  {Zhang}}]{Wu:2025ott}%
  \BibitemOpen
  \bibfield  {author} {\bibinfo {author} {\bibfnamefont {Y.}~\bibnamefont
  {Wu}}, \bibinfo {author} {\bibfnamefont {Y.}~\bibnamefont {Jiang}}, \bibinfo
  {author} {\bibfnamefont {C.}~\bibnamefont {Liu}},\ and\ \bibinfo {author}
  {\bibfnamefont {Y.}~\bibnamefont {Zhang}},\ }\bibfield  {title} {\bibinfo
  {title} {{All Giant Graviton Two-Point Functions at Two-Loops}},\ }\href@noop
  {} {\  (\bibinfo {year} {2025})},\ \Eprint {https://arxiv.org/abs/2509.23161}
  {arXiv:2509.23161 [hep-th]} \BibitemShut {NoStop}%
\bibitem [{\citenamefont {Weinberg}(2010)}]{Weinberg:2010fx}%
  \BibitemOpen
  \bibfield  {author} {\bibinfo {author} {\bibfnamefont {S.}~\bibnamefont
  {Weinberg}},\ }\bibfield  {title} {\bibinfo {title} {{Six-dimensional Methods
  for Four-dimensional Conformal Field Theories}},\ }\href
  {https://doi.org/10.1103/PhysRevD.82.045031} {\bibfield  {journal} {\bibinfo
  {journal} {Phys. Rev. D}\ }\textbf {\bibinfo {volume} {82}},\ \bibinfo
  {pages} {045031} (\bibinfo {year} {2010})},\ \Eprint
  {https://arxiv.org/abs/1006.3480} {arXiv:1006.3480 [hep-th]} \BibitemShut
  {NoStop}%
\bibitem [{\citenamefont {Simmons-Duffin}(2014)}]{Simmons-Duffin:2012juh}%
  \BibitemOpen
  \bibfield  {author} {\bibinfo {author} {\bibfnamefont {D.}~\bibnamefont
  {Simmons-Duffin}},\ }\bibfield  {title} {\bibinfo {title} {{Projectors,
  Shadows, and Conformal Blocks}},\ }\href
  {https://doi.org/10.1007/JHEP04(2014)146} {\bibfield  {journal} {\bibinfo
  {journal} {JHEP}\ }\textbf {\bibinfo {volume} {04}},\ \bibinfo {pages}
  {146}},\ \Eprint {https://arxiv.org/abs/1204.3894} {arXiv:1204.3894 [hep-th]}
  \BibitemShut {NoStop}%
\bibitem [{Note11()}]{Note11}%
  \BibitemOpen
  \bibinfo {note} {Note that our convention of $X$ and $Z$ is different from
  those in \cite {Costa:2011mg,Goncalves:2014rfa}, where $X_\protect \text
  {here}=\protect \sqrt {2}X_\protect \text {there}$ and $Z_\protect \text
  {here}=Z_\protect \text {there}/\protect \sqrt {2}$.}\BibitemShut {Stop}%
\bibitem [{\citenamefont {Costa}\ \emph {et~al.}(2016)\citenamefont {Costa},
  \citenamefont {Hansen}, \citenamefont {Penedones},\ and\ \citenamefont
  {Trevisani}}]{Costa:2016hju}%
  \BibitemOpen
  \bibfield  {author} {\bibinfo {author} {\bibfnamefont {M.~S.}\ \bibnamefont
  {Costa}}, \bibinfo {author} {\bibfnamefont {T.}~\bibnamefont {Hansen}},
  \bibinfo {author} {\bibfnamefont {J.}~\bibnamefont {Penedones}},\ and\
  \bibinfo {author} {\bibfnamefont {E.}~\bibnamefont {Trevisani}},\ }\bibfield
  {title} {\bibinfo {title} {{Projectors and seed conformal blocks for
  traceless mixed-symmetry tensors}},\ }\href
  {https://doi.org/10.1007/JHEP07(2016)018} {\bibfield  {journal} {\bibinfo
  {journal} {JHEP}\ }\textbf {\bibinfo {volume} {07}},\ \bibinfo {pages}
  {018}},\ \Eprint {https://arxiv.org/abs/1603.05551} {arXiv:1603.05551
  [hep-th]} \BibitemShut {NoStop}%
\bibitem [{Note12()}]{Note12}%
  \BibitemOpen
  \bibinfo {note} {This convention is different from \cite {Goncalves:2014rfa}
  by a factor of $2^{-J}$, which would introduce an extra $2^J$ in the
  coefficients of factorization formula comparing with \cite
  {Goncalves:2014rfa}.}\BibitemShut {Stop}%
\bibitem [{\citenamefont {Hartwell}\ and\ \citenamefont
  {Howe}(1995)}]{Hartwell:1994rp}%
  \BibitemOpen
  \bibfield  {author} {\bibinfo {author} {\bibfnamefont {G.~G.}\ \bibnamefont
  {Hartwell}}\ and\ \bibinfo {author} {\bibfnamefont {P.~S.}\ \bibnamefont
  {Howe}},\ }\bibfield  {title} {\bibinfo {title} {{(N, p, q) harmonic
  superspace}},\ }\href {https://doi.org/10.1142/S0217751X95001820} {\bibfield
  {journal} {\bibinfo  {journal} {Int. J. Mod. Phys. A}\ }\textbf {\bibinfo
  {volume} {10}},\ \bibinfo {pages} {3901} (\bibinfo {year} {1995})},\ \Eprint
  {https://arxiv.org/abs/hep-th/9412147} {arXiv:hep-th/9412147} \BibitemShut
  {NoStop}%
\end{thebibliography}%

\widetext

\begin{appendix}
\begin{center}
    \textbf{\large Supplemental Material}
\end{center}

\section{The spinning AdS$\,\times\,$S amplitudes and factorization}

In the main text, we have described the AdS$\,\times\,$S factorization formula when the exchanged operator is both a conformal scalar and a sphere scalar, such as $\mathcal O_p$. Other operators that can be exchanged (Table~\ref{tab:KK_modes}) may have conformal or sphere spin, whose factorization formula we describe in this appendix.

Conformal spinning operators can be represented in the embedding formalism~\cite{Weinberg:2010fx,Costa:2011mg,Simmons-Duffin:2012juh} by transverse sections on the projective lightcone. In our case, an operator $\mathcal X$ with conformal dimension $\Delta$ and conformal spin $J$ is encoded as
\begin{equation}
    \mathcal X(\lambda X,\alpha Z)=\lambda^{-\Delta}\alpha^J\mathcal X(X,Z),\quad\mathcal X(X,Z+\beta X)=\mathcal X(X,Z),
\end{equation}
using the (complexified) null vectors
\begin{equation}
    X^A,Z^A\in\mathbb C^{4,2},\quad X\cdot X=X\cdot Z=Z\cdot Z=0.
\end{equation}
By taking a section on the light-cone \footnote{Note that our convention of $X$ and $Z$ is different from those in \cite{Costa:2011mg,Goncalves:2014rfa}, where $X_\text{here}=\sqrt{2}X_\text{there}$ and $Z_\text{here}=Z_\text{there}/\sqrt{2}$. }
\begin{align}
    X^A = \sqrt{2}\left(x^\mu,\frac{1-x^2}{2},\frac{1+x^2}{2}\right),\qquad Z^A = \frac{1}{\sqrt{2}}\left(z^\mu,-z\cdot x , z \cdot x \right),
\end{align}
the scalar product of these null vectors reduce to invariants written in the usual 4d language
\begin{align}
    X_{ij} = -X_i \cdot X_j = x_{ij}^2,\qquad Z_i\cdot X_j = - z_i \cdot x_{ij}. 
\end{align}
On the other hand, operators in the representation $[S,R-S,S]$ of the $SU(4)$ R-symmetry can be represented as the $SO(6)$ Young diagram $[R-S,S]$ with two rows. For example, the operator $\mathcal J_5$ lives in $[1,4,1]_{SU(4)}=[4,1]_{SO(6)}$:
\begin{equation}
    \mathcal J_5=\ \raisebox{10pt}{$\yng(6,5,1)_{SU(4)}$}=\ \, \raisebox{5pt}{$\yng(5,1)_{SO(6)}$}.
\end{equation}
Following~\cite{Costa:2016hju}, we can introduce auxiliary vectors $T,W$ for each row in the $SO(6)$ Young diagram to encode the symmetry property
\begin{equation}
    \mathcal X(\lambda T,\alpha W)=\lambda^R\alpha^S\mathcal X(T,W),\quad\mathcal X(T,W+\beta T)=\mathcal X(T,W),
\end{equation}
where these null vectors satisfy
\begin{equation}
    T^A,W^A\in\mathbb C^6,\quad T\cdot T=T\cdot W=W\cdot W=0.
\end{equation}

For our purpose, the left and right sub-amplitudes always depend on at most one spinning operator $\mathcal X$, the exchanged operator. The corresponding correlator can be written as 
\begin{equation}
    G_{\mathcal X,\{p_a\}}(X_a,Z,T_a,W)=\langle \mathcal X_{p_0}(X_0,Z,T_0,W) \mathcal O_{p_1}(X_1,T_1)\cdots\mathcal O_{p_k}(X_k,T_k)\rangle,
\end{equation}
where the operators are normalized as \footnote{This convention is different from \cite{Goncalves:2014rfa} by a factor of $2^{-J}$, which would introduce an extra $2^J$ in the coefficients of factorization formula comparing with \cite{Goncalves:2014rfa}. }
\begin{equation}\label{eq:2ptnorm}
    \begin{aligned}
        \langle\mathcal X_p(X_1,Z_1,T_1,W_1)\mathcal X_p(X_2,Z_2,T_2,W_2)\rangle&=\frac{[(Z_1\cdot Z_2)(-X_1\cdot X_2)+(Z_1\cdot X_2)(Z_2\cdot X_1)]^J}{(-X_1\cdot X_2)^{\Delta+J}}\\
        &\times[(W_1\cdot W_2)(T_1\cdot T_2)-(W_1\cdot T_2)(W_2\cdot T_1)]^S(T_1\cdot T_2)^{R-S}.
    \end{aligned}
\end{equation}

We adopt the definition of the single-spin Mellin amplitudes in \cite{Goncalves:2014rfa}, and extend to the $\rm AdS\times S$ Mellin space. The corresponding amplitudes for all six master operators are defined as 
\begin{align}
    \mathcal{G}_{n+1,\mathcal{O/F}}&=\ \sumint\  \mathcal M_{n+1,\mathcal{O/F}}(\gamma_{ab},n_{ab}) \prod_{0\leq a<b\leq n}\frac{T_{ab}^{\,n_{ab}}}{n_{ab}!}\frac{\Gamma(\gamma_{ab})}{X_{ab}^{\gamma_{ab}}},\\
    \mathcal{G}_{n+1,\mathcal{J/C}}&=\sum_{c,c'=1}^n\frac{Z\cdot X_{c}}{X_{0c}} \frac{W\cdot T_{c'}}{T_{0c}}\ \sumint\ \gamma_{0c}\,n_{0c'}\,\mathcal M^{c,c'}_{n+1,\mathcal{J/C}}(\gamma_{ab},n_{ab}) \prod_{0\leq a<b\leq n}\frac{T_{ab}^{\,n_{ab}}}{n_{ab}!}\frac{\Gamma(\gamma_{ab})}{X_{ab}^{\gamma_{ab}}},\\
    \mathcal{G}_{n+1,\mathcal T}&=\sum_{c_1,c_2=1}^n\frac{Z\cdot X_{c_1}}{X_{0c_1}}\frac{Z\cdot X_{c_2}}{X_{0c_2}}\ \sumint\ \mathcal \gamma_{0c_1}(\gamma_{0c_2}+\delta_{c_1c_2})\mathcal{M}^{c_1c_2}_{n+1,\mathcal T}(\gamma_{ab},n_{ab}) \prod_{0\leq a<b\leq n}\frac{T_{ab}^{\,n_{ab}}}{n_{ab}!}\frac{\Gamma(\gamma_{ab})}{X_{ab}^{\gamma_{ab}}} ,\\
    \mathcal{G}_{n+1,\mathcal A}&=\sum_{c'_1,c'_2=1}^n\frac{W\cdot T_{c_1}}{T_{0c'_1}}\frac{W\cdot T_{c'_2}}{T_{0c'_2}}\ \sumint\ n_{0c'_1}(n_{0c'_2}-\delta_{c'_1c'_2})\mathcal{M}^{c'_1c'_2}_{n+1,\mathcal A}(\gamma_{ab},n_{ab}) \prod_{0\leq a<b\leq n}\frac{T_{ab}^{\,n_{ab}}}{n_{ab}!}\frac{\Gamma(\gamma_{ab})}{X_{ab}^{\gamma_{ab}}}.
\end{align}
These Mellin variables are symmetric in their two indices $\gamma_{ab}=\gamma_{ba},\ n_{ab}=n_{ba}$, and satisfy 
\begin{equation}
    \forall b: \quad\sum_{a\neq b} \gamma_{ab}- \sum_{a\neq b} n_{ab} = 2\delta_b .
\end{equation}
The definition of amplitudes for specific KK correlators $\mathcal{M}_{\{p_i\},\mathcal X}$ are similar, except that we should replace 
\begin{align}
    \sumint\quad \to\quad  \sumint'
\end{align}
and work on Mellin variables subject to the constraints  
\begin{equation}
    \forall b: \quad\sum_{a\neq b} \gamma_{ab}=\tau_b,\quad  \sum_{a\neq b} n_{ab} = \sigma_b.
\end{equation}

The component Mellin amplitudes are not independent but satisfy ``gauge invariance'' conditions. For $\mathcal J$ and $\mathcal C$ the conditions are 
\begin{equation}
    \sum_{c=1}^n\gamma_{0c}\mathcal M_{\mathcal J/\mathcal C}^{c,c'}=0,\qquad\sum_{c'=1}^nn_{0c'}\mathcal M_{\mathcal J/\mathcal C}^{c,c'}=0,
\end{equation}
and for $\mathcal T$ and $\mathcal A$,
\begin{equation}
    \sum_{c_1=1}^n(\gamma_{0c_1}+\delta_{c_1c_2})\mathcal M_{\mathcal T}^{c_1c_2}=0\qquad \sum_{c'_1=1}^n(n_{0c'_1}-\delta_{c'_1c'_2})\mathcal M_{\mathcal A}^{c'_1c'_2}=0. 
\end{equation}
Under $Z_2$ symmetry, these amplitudes transform as 
\begin{align}
    \mathcal{M}_{n,\mathcal{O}}(\gamma_{ij},n_{ij}) \to&\  (-)^n \mathcal{M}_{n,\mathcal{O}}(-n_{ij},-\gamma_{ij}),\\
    \mathcal{M}_{n,\mathcal{J}}^{c,c'}(\gamma_{ij},n_{ij}) \to&\ (-)^n \mathcal{M}_{n,\mathcal{J}}^{c',c}(-n_{ij},-\gamma_{ij}), \\ \mathcal{M}_{n,\mathcal{T}}^{c_1c_2}(\gamma_{ij},n_{ij}) \to&\ (-)^{n+1} \mathcal{M}_{n,\mathcal{A}}^{c'_1c'_2}(-n_{ij},-\gamma_{ij}), \\
    \mathcal{M}_{n,\mathcal{A}}^{c'_1c'_2}(\gamma_{ij},n_{ij}) \to&\ (-)^{n+1} \mathcal{M}_{n,\mathcal{T}}^{c_1c_2}(-n_{ij},-\gamma_{ij}), \\
    \mathcal{M}_{n,\mathcal{C}}^{c,c'}(\gamma_{ij},n_{ij}) \to&\ (-)^{n+1} \mathcal{M}_{n,\mathcal{C}}^{c',c}(-n_{ij},-\gamma_{ij}), \\
    \mathcal{M}_{n,\mathcal{F}}(\gamma_{ij},n_{ij}) \to&\  (-)^n \mathcal{M}_{n,\mathcal{F}}(-n_{ij},-\gamma_{ij}).
\end{align}

The Mellin factorization formula for the exchange of operators with conformal spin $\leq2$ has been worked out in~\cite{Goncalves:2014rfa}, which we copy below for the specific case of $d=4$, in terms of the $\mhat{\gamma}$ symbols. The exchange of a conformal scalar with dimension $\Delta$ corresponds to poles
\begin{equation}
    M(\gamma_{ab})\sim\frac{Q_m}{\gamma_{LR}-(\tau+2m)},\quad Q_m=\frac{-2\Gamma(\tau)m!}{(\tau-1)_m}M_{L,m}M_{R,m},
\end{equation}
where
\begin{equation}
    M_{L,m}=\frac1{m!}(\widehat\gamma_{LL})^m\circ M_L,\quad M_{R,m}=\frac1{m!}(\widehat\gamma_{RR})^m\circ M_R,
\end{equation}
and
\begin{equation}
    \widehat\gamma_{LL}=\sum_{a,b\in L,a<b}\widehat\gamma_{ab},\quad\widehat\gamma_{ab}\circ M(\gamma_{ab})=\gamma_{ab}M(\gamma_{ab}+1),\quad\text{similarly for }\widehat\gamma_{RR}.
\end{equation}
The exchange of a conformal spin-1 operator with dimension $\Delta$ corresponds to poles (we sum over $a,b\in L$ and $i,j\in R$ when these indices appear twice in the expressions)
\begin{equation}
    M(\gamma_{ab})\sim\frac{Q_m}{\gamma_{LR}-(\tau+2m)},\quad Q_m=\frac{2(\tau+1)\Gamma(\tau)m!}{(\tau)_m}\left(\widehat\gamma_{ai}-\frac{\tau-1}{m(\tau-2)}\widehat\gamma_{aR}\widehat\gamma_{iL}\right)\circ M_{L,m}^aM_{R,m}^i,
\end{equation}
where
\begin{equation}
    \widehat\gamma_{aR}=\sum_{i\in R}\widehat\gamma_{ai},\quad\widehat\gamma_{iL}=\sum_{a\in L}\widehat\gamma_{ai}.
\end{equation}
The exchange of a conformal spin-2 operator with dimension $\Delta$ corresponds to poles 
\begin{equation}
    M(\gamma_{ab})\sim\frac{Q_m}{\gamma_{LR}-(\tau+2m)},\quad Q_m=h_m^{(1)}Q_m^{(1)}+\cdots+h_m^{(5)}Q_m^{(5)},
\end{equation}
where
\begin{equation}
    \begin{gathered}
        h_m^{(1)}=\frac{-2(\tau+3)\Gamma(\tau+1)m!}{(\tau+1)_m},\quad h_m^{(2)}=\frac{-2\tau}{m(\tau-2)}h_m^{(1)},\quad h_m^{(3)}=\frac{-(\tau+1)}{2(m-1)(\tau-1)}h_m^{(2)},\\
        h_m^{(4)}=\frac{-(\tau+m)}{\tau+1}h_m^{(3)},\quad h_m^{(5)}=-\left(1+2(m-1)\left(1-\frac{(\tau+1)(\tau+2)}{8\tau}\right)\right)h_m^{(4)},
    \end{gathered}
\end{equation}
and
\begin{equation}
\begin{aligned}
    Q_m^{(1)}&=\widehat\gamma_{ai}\widehat\gamma_{bj}\circ M_{L,m}^{ab}M_{R,m}^{ij},\\
    Q_m^{(2)}&=\widehat\gamma_{ai}\widehat\gamma_{bR}\widehat\gamma_{jL}\circ M_{L,m}^{ab}M_{R,m}^{ij},\\
    Q_m^{(3)}&=\widehat\gamma_{aR}\widehat\gamma_{bR}\widehat\gamma_{iL}\widehat\gamma_{jL}\circ M_{L,m}^{ab}M_{R,m}^{ij},\\
    Q_m^{(4)}&=\widehat\gamma_{ab}\widehat\gamma_{iL}\widehat\gamma_{jL}\circ M_{L,m-1}^{ab}M_{R,m}^{ij}+\widehat\gamma_{aR}\widehat\gamma_{bR}\widehat\gamma_{ij}\circ M_{L,m}^{ab}M_{R,m-1}^{ij},\\
    Q_m^{(5)}&=\widehat\gamma_{ab}\widehat\gamma_{ij}\circ M_{L,m-1}^{ab}M_{R,m-1}^{ij}.
\end{aligned}
\end{equation}
The spherical factorization formula can be obtained by performing the Wick rotation on the above expressions
\begin{align}
    (\Delta,J,\tau,m,\gamma_{ij})\leftrightarrow (-R,S,-\sigma,r,-n_{ij})\,,\qquad M\big|_{\gamma_{ij}\to \gamma_{ij}\pm 1}\leftrightarrow M\big|_{n_{ij}\to n_{ij}\mp 1}\,,
\end{align}
and multiplying an extra factor on the formula 
\begin{align}
    -\frac{1}{2R!\,\Gamma(-R)}
\end{align}
to make the normalization correct. This can also be checked for explicit examples using the differential operators in~\cite{Costa:2016hju}. Moreover, the factorization is not on poles, but simply at specific values in the amplitudes
\begin{align}
    M(\gamma_{ab})\sim\frac{Q_m}{\gamma_{LR}-(\tau+2m)}\ \leftrightarrow\ M(n_{ab})\Big|_{n_{LR}=\sigma-2r} = Q_r
\end{align}

Combining the AdS part and the sphere part of the factorization as a direct product, we easily obtain the AdS$\,\times\,$S factorization formula corresponding to the exchange of an operator $\mathcal X_p$ with definite quantum numbers $(\Delta,J|R,S)$. Concretely, this corresponds to poles of the form
\begin{equation}
    \mathcal M_{\{p_i\}}\Big|_{n_{LR}=\sigma-2r}\sim\frac{Q_{m,r}}{\gamma_{LR}-(\tau+2m)}\Big|_{n_{LR}=\sigma-2r},
\end{equation}
where $Q_{m,r}$ denotes the combined action (shifting and gluing) in $Q_m$ and $Q_r$ on both the AdS and the sphere Mellin variables. For example, for the exchange of $\mathcal J_p$
\begin{equation}
    \begin{aligned}
        Q_{m,r}=&\ \frac{2(\tau+1)\Gamma(\tau)m!}{(\tau)_m}\!\left(\widehat\gamma_{ai}-\frac{\tau-1}{m(\tau-2)}\widehat\gamma_{aR}\widehat\gamma_{iL}\!\right)\circ \\
        &\quad \frac{(-1)^{r+1} (\sigma-1)r!}{[\sigma]_r\sigma!}\left(-\widehat n_{a'i'}-\frac{\sigma+1}{r(\sigma+2)}\widehat n_{a'R}\widehat n_{i'L}\!\right)\circ \left(\mathcal{M}_{L,\{p_i\},\mathcal{J}_p}^{a,a'}\!\right)_{m,r}\!\left(\mathcal{M}_{R,\{p_i\},\mathcal{J}_p}^{i,i'}\!\right)_{m,r}\!,
    \end{aligned}
\end{equation}
with
\begin{align}
    \left(\mathcal{M}_{L,\{p_i\},\mathcal{J}_p}^{a,a'}\!\right)_{m,r} = \frac{\left(\mhat{\gamma}_{LL}\right)^m}{m!}\frac{\left(\mhat{n}_{LL}\right)^r}{r!} \circ \mathcal{M}_{L,\{p_i\},\mathcal{J}_p}^{a,a'},\quad \text{similarly for $\mathcal{M}_{R,\{p_i\},\mathcal{J}_p}^{i,i'}$.}
\end{align}
Crucially, by looking at Table~\ref{tab:KK_modes} of quantum numbers, we see that $2\delta\equiv\tau-\sigma$ depends on the type $\mathcal X$ but not the Kaluza-Klein level $p$. Since the residue is only defined on the support of $\gamma_{LR}=\tau+2m$ and $n_{LR}=\sigma-2r$, we can rewrite the formula as 
\begin{equation}\label{eq:adsxsder}
    \mathcal M_{\{p_i\}}\Big|_{n_{LR}=\sigma-2r}\sim \frac{Q_{m,r}}{\gamma_{LR}-(\tau+2m)-(n_{LR}-\sigma+2r)}\Big|_{n_{LR}=\sigma-2r} =\frac{\mathcal{Q}_k}{\rho_{LR}-(2\delta+2k)}\Big|_{n_{LR}=\sigma-2r},
\end{equation}
with $\rho_{LR}=\gamma_{LR}-n_{LR}$, $k=m+r$ and $\mathcal{Q}_k = \sum_{m+r=k}Q_{m,r}$.

At this stage, we are still discussing the factorization on an operator $\mathcal{X}_p$ with a specific KK level $p$ in the correlator $G_{\{p_i\}}$. However, the true power of AdS$\,\times\,$S formalism is the ability to manipulate expressions with ease at the level of generating functions. In \eqref{eq:adsxsder}, when both sub-amplitudes are analytic in $p_i$, we can replace $p_i$ with Mellin variables, and the same amplitude now describe the generating function of the whole family of correlators. Similarly, on the support, we can replace $\tau\to\gamma_{LR}-2m$ and $\sigma\to n_{LR}+2r$ without changing the residue. The resulting formula is independent of level $p$, thus valid for all KK levels simultaneously. This can also be understood as we are interpolating the $p_i$ dependence among different KK correlators, and the $p$ dependence among the factorization of different multiplets, using the Mellin variables $\gamma_{ab}$ and $n_{ab}$. Therefore, it is not surprising that the $\rm AdS\times S$ factorization formula can work for arbitrary KK level. 

Sometimes it would be more convenient not to normalize the operators by \eqref{eq:2ptnorm}, but introduce an extra normalization factor $\mathcal{N}_\mathcal{X}$ (which might depends on $p$)
\begin{equation}
    \begin{aligned}
        \langle\mathcal X_p(X_1,Z_1,T_1,W_1)\mathcal X_p(X_2,Z_2,T_2,W_2)\rangle&=\mathcal{N}_{\mathcal{X}} \frac{[(Z_1\cdot Z_2)(-X_1\cdot X_2)+(Z_1\cdot X_2)(Z_2\cdot X_1)]^J}{(-X_1\cdot X_2)^{\Delta+J}}\\
        &\times[(W_1\cdot W_2)(T_1\cdot T_2)-(W_1\cdot T_2)(W_2\cdot T_1)]^S(T_1\cdot T_2)^{R-S}.
    \end{aligned}
\end{equation}
The master operators $\mathcal X=\sum_p\mathcal X_p$ we use are always defined with respect to this new normalization. In switching to the new normalization, the $n$-point function $\langle \mathcal{X}\mathcal{O} \cdots \mathcal{O} \rangle$ in AdS$\,\times\,$S Mellin space is renormalized as
\begin{align}
    \mathcal{M}_{n,\mathcal{X}} \to \sqrt{\mathcal{N}_\mathcal{X}(p_0)\prod_{i=1}^{n-1}\mathcal{N}_\mathcal{O}(p_i)}\,  \mathcal{M}_{n,\mathcal{X}},
\end{align}
and introduce an additional factor in the coefficient of the factorization formula
\begin{align}
    \mathcal{Q}_k \to \frac{1}{\mathcal{N}_\mathcal{X}}\mathcal{Q}_k
\end{align}
when we consider the exchange of $\mathcal{X}$. One can use this normalization factor to cancel the unwanted factors in the amplitudes, when these unwanted factors factorized as functions of external KK levels. The normalization factors we adopt in our computation are
\begin{align}
    \mathcal N_{\mathcal O}&=\frac{1}{p},\\
    \mathcal N_{\mathcal J}&=(p-1)p(p+1),\\
    \mathcal N_{\mathcal T}&=(p-1)p(p+1)^2(p+2)^2(p+3),\\
    \mathcal N_{\mathcal A}&=(p-3)(p-2)^2(p-1)^2p(p+1),\\
    \mathcal N_{\mathcal C}&=(p-3)(p-2)^2(p-1)^2p^3(p+1)^2(p+2)^2(p+3),\\
    \mathcal N_{\mathcal F}&=(p-3)(p-2)^2(p-1)^3p^3(p+1)^3(p+2)^2(p+3).
\end{align}
Note that these $\mathcal{N}_\mathcal{X}$ also respect the $Z_2$ symmetry. Under such convention, the three-point superdescendant amplitudes $\mathcal{M}_{3,\mathcal{X}}$ obtained from the factorization of four-point function can be recast into polynomials, which we summarize the results here (with $I$ the label of $\mathcal X$)
\begin{align}
    \mathcal{M}_{3,\mathcal{O}} =&\ n_{12}n_{1I}n_{2I}, \\
    \mathcal{M}^{c,c'}_{3,\mathcal{J}} =&\ 
    \begin{pmatrix}
        \ n_{12} \gamma_{2I}n_{2I} & -n_{12} \gamma_{2I}n_{1I} \\
        -n_{12} \gamma_{1I}n_{2I} & \quad n_{12} \gamma_{1I}n_{1I}
    \end{pmatrix}, \\
    \mathcal{M}^{c_1c_2}_{3,\mathcal{T}} =&\
    \begin{pmatrix}
        n_{12}(n_{12}-1) \gamma_{2I}(\gamma_{2I}+1) &  -n_{12}(n_{12}-1) (\gamma_{2I}+1)(\gamma_{1I}+1) \\
        -n_{12}(n_{12}-1) (\gamma_{1I}+1)(\gamma_{2I}+1) & n_{12}(n_{12}-1) \gamma_{1I}(\gamma_{1I}+1)
    \end{pmatrix}, \\
    \mathcal{M}^{c'_1c'_2}_{3,\mathcal{A}} =&\
    \begin{pmatrix}
        n_{12}(n_{12}-1) n_{2I}(n_{2I}-1) & -n_{12}(n_{12}-1) (n_{2I}-1)(n_{1I}-1) \\
        -n_{12}(n_{12}-1) (n_{1I}-1)(n_{2I}-1) & n_{12}(n_{12}-1) n_{1I}(n_{1I}-1)
    \end{pmatrix}, \\
    \mathcal{M}^{c,c'}_{3,\mathcal{C}} =&\
    (n_{1I}+n_{2I}+2)\begin{pmatrix}
        n_{12}(n_{12}-1)(n_{12}-2) \gamma_{2I}n_{2I} & -n_{12}(n_{12}-1)(n_{12}-2) \gamma_{2I}n_{1I} \\
        -n_{12}(n_{12}-1)(n_{12}-2) \gamma_{1I}n_{2I} & n_{12}(n_{12}-1)(n_{12}-2) \gamma_{1I}n_{1I}
    \end{pmatrix}, \\
    M_{3,\mathcal{F}} =&\ n_{12}(n_{12}-1)(n_{12}-2)(n_{12}-3)(n_{12}-4)(n_{1I}+2)(n_{2I}+2).
\end{align}

\section{Drukker-Plefka twist in $\rm AdS\times S$ space}

For a generic KK correlator, performing the DP twist sends the function to a constant 
\begin{align*}
    G_{\{p_i\}}(X_{ij},T_{ij})\big|_{T_{ij}\to X_{ij}} = \text{const.}
\end{align*}
The same twist works for the generating function $\mathcal{G}$. However, if we simply take $T_{ij}\to X_{ij}$ in the generating function, functions with different KK weights would mix together, and give rise to a weaker constraint. In order to track different KK correlators, we can first take $T_{ij}\to a_i a_j T_{ij}$, where each correlator acquires the weight $\prod_i a_i^{p_i}$, and then safely do the DP twist $T_{ij}\to X_{ij}$. In $\rm AdS\times S$ space, this gives 
\begin{align}\label{eq:AdSxSDP}
    \mathcal{G}\big|_{\text{DP twist}} =&\ \sumint\ \mathcal{M}(\gamma_{ij},n_{ij}) \prod_{i<j} \frac{\Gamma(\gamma_{ij})}{n_{ij}!} \frac{(a_ia_j)^{n_{ij}}}{X_{ij}^{\gamma_{ij}-n_{ij}}}= \sumint\ \mathcal{M}(\rho_{ij}+n_{ij},n_{ij}) \prod_{i<j} \frac{\Gamma(\rho_{ij}+n_{ij})}{n_{ij}!} \frac{(a_ia_j)^{n_{ij}}}{X_{ij}^{\rho_{ij}}}.
\end{align}
The amplitude $\mathcal{M}$, when written in terms of $\rho_{ij}$ and $n_{ij}$, is a polynomial of $n_{ij}$. Note that, since $[n]_c=n(n-1)\cdots(n-c+1)$ form a basis for polynomials of $n$, we can always decompose the amplitude as
\begin{align*}
    \mathcal{M}(\rho_{ij}+n_{ij},n_{ij}) \equiv \sum_{k} \mathcal{M}^k(\rho_{ij}) \prod_{i<j}\,[n_{ij}]_{c_{ij,k}}\,,
\end{align*}
where $c_{ij,k}$ are integers. In this decomposition, we can perform the sum over $n_{ij}$ in \eqref{eq:AdSxSDP} for each $k$, which gives the integrand
\begin{align}
    \sum_k \mathcal M^k(\rho_{ij}) \prod_{i<j} \frac{}{}(1-a_ia_j)^{-\rho_{ij}-c_{ij,k}}(a_i a_j)^{c_{ij,k}}\Gamma(\rho_{ij}+c_{ij,k})X_{ij}^{-\rho_{ij}}.
\end{align}
Therefore the result can be recast into a Mellin integral on $\rho$
\begin{align}
    \mathcal{G}\big|_{\text{DP twist}} = \int \mathcal{I}\,\prod_{i<j}\frac{\Gamma(\rho_{ij})}{\left((1-a_ia_j)X_{ij}\right)^{\rho_{ij}}},
\end{align}
where 
\begin{align} 
    \mathcal{I} \equiv \sum_k \mathcal{M}^k(\rho_{ij}) \prod_{i<j}\left(\frac{a_i a_j}{1-a_ia_j}\right)^{c_{ij,k}} (\rho_{ij})_{c_{ij,k}}.
\end{align}
Since the Mellin amplitude of a constant should be 0, we require $\mathcal{I}=0$. This gives the $\rm AdS\times S$ DP twist condition in the main text. 

\section{The expression of $P^{(0,0)}(\gamma_{ij},n_{ij})$}

We present the final result of the double pole residue $P^{(0,0)}(\gamma_{ij},n_{ij})$ here
\begin{equation}
    \begin{aligned}
        P^{(0,0)}(\gamma_{ij},n_{ij}) =&\ n_{12} \left(n_{13}\!+\!n_{14}\!+\!n_{15}\!+\!n_{23}\!+\!n_{24}\!+\!n_{25}\right)\left(n_{14}\!+\!n_{15}\!+\!n_{24}\!+\!n_{25}\!+\!n_{34}\!+\!n_{35}\right) n_{45} \\
        & \left(\!-\!\gamma _{15}\left(n_{13}\!+\!n_{23}\right) \left(n_{15} n_{24}\!-\!n_{14} n_{25}\right)\left(n_{34}\!+\!n_{35}\right)\!-\!\gamma _{34} \left(n_{13}\!+\!n_{23}\right) \left(n_{15}\left(\left(n_{23}\!+\!n_{25}\right) n_{34}\!-\!n_{24} n_{35}\right) \right.\right. \\
        & \left. \left. \!+\!n_{14} \left(n_{25} n_{34}\!-\!\left(n_{23}\!+\!n_{24}\right) n_{35}\right)\right)\!+\!\gamma _{23} \left(\gamma _{34} \left(n_{15} n_{23} n_{34}\!-\!n_{14} n_{23} n_{35}\!+\!n_{13} \left(n_{24} n_{35}\!-\!n_{25} n_{34}\right)\right) \right.\right.\\
        & \left.\left.\!-\!\left(\left(n_{14} n_{23}\!-\!n_{13} n_{24}\right)\left(n_{15}\!+\!n_{25}\right)\!+\!\left(n_{15} n_{23}\!-\!n_{13} n_{25}\right) n_{34}\right)\left(n_{34}\!+\!n_{35}\right)\right) \right.\\
        & \left. \!+\!\left(n_{13}\!+\!n_{23}\right) \left(\left(n_{15} n_{24}\!\!+\!\!n_{14}\left(n_{23}\!\!+\!\!n_{24}\right)\right) \left(n_{15}\!\!+\!\!n_{25}\right)\!+\!\left(n_{14} n_{25}\!+\!n_{15}\left(n_{23}\!+\!n_{25}\right)\right) n_{34}\right) \left(n_{34}\!+\!n_{35}\right)\right).
    \end{aligned}
\end{equation}
The full result of $\mathcal M_5$ can be found in the ancillary file.

\section{Superspace method}

$\mathcal{N}=4$ harmonic/analytic superspace \cite{Hartwell:1994rp} linearly realizes $PSL(2|2)\times PSL(2|2)\times GL(1)^2\subset PSL(4|4)$ of the (complexified) symmetry group. Specifically, the (complexified) superspace is equipped with coordinates:
\begin{equation}
    \left(\begin{matrix}
        x_{\alpha\dot\beta} & \varrho_{\alpha\dot b} \\ \lambda_{a\dot\beta} & y_{a\dot b}
    \end{matrix}\right),
\end{equation}
where $x$ denotes the ordinary spacetime position, $y$ denotes the R-symmetry polarization, and $\varrho,\lambda$ are Grassmann variables on the supermanifold. A superfield contains all the fields in a given supermultiplet through the expansion in the Grassmann variables. For example, the half-BPS supermultiplets are encoded in the scalar superfields, which reads schematically:
\begin{equation}
    \mathbb O_p(x,\varrho,\lambda,y)=\mathcal O_p(x,y)+\varrho\lambda\mathcal J_p(x,y)+\varrho^2\lambda^2\mathcal T_p(x,y)+\varrho^2\lambda^2\mathcal A_p(x,y)+\varrho^3\lambda^3\mathcal C_p(x,y)+\varrho^4\lambda^4\mathcal F_p(x,y)+\cdots
\end{equation}
where the dots contain other (chiral or fermionic) superprimary operators which we do not need. One can isolate a specific operator in the supermultiplet by acting on the superfield with a differential operator and then setting Grassmann variables to zero (up to an overall normalization which we would not care too much in this section)
\begin{equation}
    \mathcal X_p(x,y) = \mathcal{D}^{(\mathcal X_p)}\mathbb O_p(x,\varrho,\lambda,y)\Big|_{\varrho=\lambda=0}.
\end{equation}
For example:
\begin{align*}
    &\qquad\qquad\qquad \mathcal{D}^{(\mathcal O_p)}=\, 1,\qquad\qquad\qquad\qquad\mathcal{D}^{(\mathcal J_p)} = \frac{\partial}{\partial \lambda^{a\dot\alpha}} \frac{\partial}{\partial \varrho^{\alpha\dot a}} - \frac{1}{p} \frac{\partial}{\partial x^{\alpha \dot{\alpha}}} \frac{\partial}{\partial y^{a\dot{a}}}, \\
    \mathcal{D}^{(\mathcal{T}_p)} =\, & \varepsilon^{ab}\varepsilon^{\dot{a}\dot{b}} \left( \frac{\partial}{\partial\lambda^{b\dot{\beta}}}\frac{\partial}{\partial\lambda^{a \dot{\alpha}}}\frac{\partial}{\partial\varrho^{\beta\dot{b}}}\frac{\partial}{\partial\varrho^{\alpha\dot{a}}}-\frac{4}{p+2}\frac{\partial}{\partial\lambda^{b\color{blue}{(\dot{\beta}}}}\frac{\partial}{\partial x^{\color{red}{(\alpha}\color{black}|\color{blue}{\dot{\alpha})}}}\frac{\partial}{\partial y^{a \dot{a}}}\frac{\partial}{\partial\varrho^{\color{red}{\beta)}\color{black} \dot{b}}}-\frac{2}{(p+1)(p+2)}\frac{\partial}{\partial x^{\color{olive}(\alpha\color{black}| \dot \alpha}}\frac{\partial}{\partial x^{\color{olive}\beta)\color{black} \dot \beta }}\frac{\partial}{\partial y^{a \dot a}}\frac{\partial}{\partial y^{b \dot b}}\right), \\
    \mathcal{D}^{(\mathcal{A}_p)} =\,  &\varepsilon^{\alpha\beta}\varepsilon^{\dot{\alpha}\dot{\beta}} \left( \frac{\partial}{\partial\lambda^{b\dot{\beta}}}\frac{\partial}{\partial\lambda^{a \dot{\alpha}}}\frac{\partial}{\partial\varrho^{\beta\dot{b}}}\frac{\partial}{\partial\varrho^{\alpha\dot{a}}}-\frac{4}{p-2}\frac{\partial}{\partial\lambda^{\color{blue}(b\color{black}|\dot{\beta}}}\frac{\partial}{\partial x^{\alpha\dot{\alpha}}}\frac{\partial}{\partial y^{\color{blue}a)\color{red}( \dot{a}}}\frac{\partial}{\partial\varrho^{\beta|\color{red}\dot{b})}}-\frac{2}{(p-1)(p-2)}\frac{\partial}{\partial x^{\alpha \dot \alpha}}\frac{\partial}{\partial x^{\beta \dot \beta }}\frac{\partial}{\partial y^{\color{olive}(a\color{black}| \dot a}}\frac{\partial}{\partial y^{\color{olive}b)\color{black} \dot b}}\right).
\end{align*}
where curved brackets denote index symmetrization. The $Z_2$ symmetry acts on these differential operators by mapping $p\leftrightarrow -p$ as well as swapping Greek and Latin indices: $x\leftrightarrow y,\ \varrho\leftrightarrow\lambda$. As a result, $\mathcal D^{(\mathcal J_p)}\leftrightarrow-\mathcal D^{(\mathcal J_p)}$ and $\mathcal D^{(\mathcal T_p)}\leftrightarrow\mathcal D^{(\mathcal A_p)}$. Four-point functions of the form
\begin{equation*}
    \langle\mathcal X_{p_1}(x_1,y_1)\mathcal O_{p_2}(x_2,y_2)\mathcal O_{p_3}(x_3,y_3)\mathcal O_{p_4}(x_4,y_4)\rangle
\end{equation*}
can be obtained by acting the differential operators on the four-point function of superfields:
\begin{equation}
    \langle\mathcal X_{p_1}\mathcal O_{p_2}\mathcal O_{p_3}\mathcal O_{p_4} \rangle =  \mathcal{D}_1^{(\mathcal X_{p_1})} \langle \mathbb{O}_{p_1}\mathbb{O}_{p_2}\mathbb{O}_{p_3}\mathbb{O}_{p_4} \rangle \Big|_{\varrho_i=\lambda_i=0}.
\end{equation}
The four-point function of superfields can, in turn, be obtained by uplifting the four-point function $\langle\mathcal O_{p_1}\mathcal O_{p_2}\mathcal O_{p_3}\mathcal O_{p_4}\rangle$ to superspace, using the procedure detailed in \cite{Goncalves:2023oyx}.

The differential operators can be determined as follows. First, we can make a generic ansatz compatible with the charges and symmetries of each operator. For example, for $\mathcal{J}_p$ the ansatz would take the form
\begin{equation}
    \mathcal{D}^{(\mathcal J_p)} = \frac{\partial}{\partial \lambda^{a\dot\alpha}} \frac{\partial}{\partial \varrho^{\alpha\dot a}} + \mu \frac{\partial}{\partial x^{\alpha \dot{\alpha}}} \frac{\partial}{\partial y^{a\dot{a}}}.
\end{equation}
up to an overall constant which amounts to a choice of normalization. We can then consider the two-point function
\begin{equation}
    \langle \mathcal J_p(x_1,y_1)\mathcal O_p(x_2,y_2)\rangle = \mathcal{D}^{(\mathcal J_p)}_1 \langle \mathbb{O}_p(x_1,y_1)\mathbb{O}_p(x_2,y_2)\rangle \Big|_{\varrho_i=\lambda_i=0},
\end{equation}
and derive a constraint on $\mu$ by plugging in our ansatz for $\mathcal D^{(\mathcal{J}_p)}$ and imposing that the two-point function is diagonal, $\langle \mathcal J_p \mathcal O_p\rangle = 0$. An analogous procedure can be done for the other superdescendants. Since $\langle\mathbb O_p(x_1,y_1)\mathbb O_p(x_2,y_2)\rangle=\frac1p(\frac{y_{12}^2}{x_{12}^2})^p+\cdots$ transforms nicely under the $Z_2$ symmetry, the above procedure naturally yields differential operators with $Z_2$ symmetry. In this way, using the input from the known results for the four-point function $\langle\mathcal O_{p_1}\mathcal O_{p_2}\mathcal O_{p_3}\mathcal O_{p_4} \rangle$, we explicitly obtained
\begin{equation}
    \langle\mathcal J_{p_1}\mathcal O_{p_2}\mathcal O_{p_3}\mathcal O_{p_4} \rangle,\quad \langle \mathcal T_{p_1}\mathcal O_{p_2}\mathcal O_{p_3}\mathcal O_{p_4} \rangle,\quad \langle \mathcal A_{p_1}\mathcal O_{p_2}\mathcal O_{p_3}\mathcal O_{p_4} \rangle.
\end{equation}
These results can be compared with the results from the bootstrap computation, and they completely match up to an overall normalization.

One could, in principle, carry out the same procedure for the four-point functions involving $\mathcal C_p$ and $\mathcal F_p$. This is, however, very computationally demanding, since these operators are higher-level superdescendants of $\mathcal O_p$ and thus require more derivatives in their differential operators. For this reason, we were unable to compute the remaining four-point functions using the superspace method explicitly.

\section{Computation of R-factors}

The possible R-factors for multi-point correlation functions are computed following the method reviewed in~\cite{Heslop:2022xgp}, which we summarize here briefly.

The correlator $\langle\mathcal O\mathcal O\mathcal O\mathcal O\mathcal O\rangle$ that we study is the $\varrho=\lambda=0$ component of $\langle\mathbb O\mathbb O\mathbb O\mathbb O\mathbb O\rangle$. For this reason, we may restrict to the \emph{chiral} subspace $\lambda=0$. Within the chiral subspace, under a chiral superconformal transformation ${\Xi_{\alpha}}^B{Q^\alpha}_B+\Xi^{\dot\alpha B}\bar S_{\dot\alpha B}$, the $x,y$ coordinates are unchanged while $\varrho$ is translated by
\begin{equation}
    \varrho_{\alpha\dot b}\mapsto\hat\varrho_{\alpha\dot b}(\Xi)=\varrho_{\alpha\dot b}+{x_\alpha}^{\rm A}{\Xi_{\rm A}}^By^\perp_{B\dot b},
\end{equation}
where
\begin{equation}
    {x_\alpha}^{\rm B}=\left(\begin{matrix}
        {\delta_\alpha}^\beta & x_{\alpha\dot\beta}
    \end{matrix}\right),\quad y^\perp_{A\dot b}=\left(\begin{matrix}
        -y_{a\dot b} \\[2mm] {\delta^{\dot a}}_{\dot b}
    \end{matrix}\right),
\end{equation}
and ${\Xi_{\rm A}}^B=({\Xi_\alpha}^B,\Xi^{\dot\alpha B})$ parametrizes the transformation. As a result, functions at $\varrho=\lambda=0$ and invariant under $Q,\bar S$ can be constructed by integrating over the chiral subspace. Since we are dealing with 5-point correlators, we only need
\begin{equation}
    \mathcal I\left(\begin{smallmatrix}
        {\rm A_1} & {\rm A_2} & {\rm A_3} & {\rm A_4} \\ i_1 & i_2 & i_3 & i_4 \\ B_1 & B_2 & B_3 & B_4
    \end{smallmatrix}\right)=\int{\rm d}^{16}\Xi(y^\perp\partial_{\hat\varrho}x)_{i_1B_1}^{{\rm A}_1}(y^\perp\partial_{\hat\varrho}x)_{i_2B_2}^{{\rm A}_2}(y^\perp\partial_{\hat\varrho}x)_{i_3B_3}^{{\rm A}_3}(y^\perp\partial_{\hat\varrho}x)_{i_4B_4}^{{\rm A}_4}(\hat\varrho_1^4\hat\varrho_2^4\hat\varrho_3^4\hat\varrho_4^4\hat\varrho_5^4).
\end{equation}
The R-factors we seek are scalar functions obtained by contracting the above invariant with $x_{\rm AB},\epsilon_{{\rm ABCD}},y^{AB},\epsilon^{ABCD}$, where
\begin{equation}
    x_{\rm AB}=\frac12\epsilon_{\rm ABCD}{x_\alpha}^{\rm C}{x_\beta}^{\rm D}\epsilon^{\alpha\beta},\quad y^{AB}=\frac12\epsilon^{ABCD}y^\perp_{C\dot a}y^\perp_{D\dot b}\epsilon^{\dot a\dot b}.
\end{equation}
By construction, the R-factors are invariant under $Q,\bar S$. Moreover, since it is a scalar function depending on $(X_i\cdot X_j)\sim x_{ij}^2$ and $(T_i\cdot T_j)\sim y_{ij}^2$, it is parity even. Together these imply that the R-factors are really invariant (more precisely, covariant, because we made no effort in adjusting the conformal weights) under the full $\mathcal N=4$ superconformal symmetry.

Depending on whether $\epsilon$ or $x,y$ is used to contract the indices, we divide the resulting R-factors into 5 classes:
\begin{align*}
    \mathcal I^{\epsilon\epsilon}(i_1i_2i_3i_4)&=\epsilon_{{\rm ABCD}}\epsilon^{ABCD}\mathcal I\left(\begin{smallmatrix}
        {\rm A} & {\rm B} & {\rm C} & {\rm D} \\ i_1 & i_2 & i_3 & i_4 \\ A & B & C & D
    \end{smallmatrix}\right)\sim(X\cdot X)^5(T\cdot T)^5;\\
    \mathcal I^{x\epsilon}(j_1j_2;i_1i_2i_3i_4)&=x^{j_1}_{\rm AB}x^{j_2}_{\rm CD}\epsilon^{ABCD}\mathcal I\left(\begin{smallmatrix}
        {\rm A} & {\rm B} & {\rm C} & {\rm D} \\ i_1 & i_2 & i_3 & i_4 \\ A & B & C & D
    \end{smallmatrix}\right)\sim(X\cdot X)^6(T\cdot T)^5;\\
    \mathcal I^{\epsilon y}(k_1k_2;i_1i_2i_3i_4)&=\epsilon_{{\rm ABCD}}y_{k_1}^{AB}y_{k_2}^{CD}\mathcal I\left(\begin{smallmatrix}
        {\rm A} & {\rm B} & {\rm C} & {\rm D} \\ i_1 & i_2 & i_3 & i_4 \\ A & B & C & D
    \end{smallmatrix}\right)\sim(X\cdot X)^5(T\cdot T)^6;\\
    \mathcal I^{xy}_{\rm I}(j_1j_2;k_1k_2;i_1i_2i_3i_4)&=x^{j_1}_{\rm AB}x^{j_2}_{\rm CD}y_{k_1}^{AB}y_{k_2}^{CD}\mathcal I\left(\begin{smallmatrix}
        {\rm A} & {\rm B} & {\rm C} & {\rm D} \\ i_1 & i_2 & i_3 & i_4 \\ A & B & C & D
    \end{smallmatrix}\right)\sim(X\cdot X)^6(T\cdot T)^6;\\
    \mathcal I^{xy}_{\rm II}(j_1j_2;k_1k_2;i_1i_2i_3i_4)&=x^{j_1}_{\rm AB}x^{j_2}_{\rm CD}y_{k_1}^{AC}y_{k_2}^{BD}\mathcal I\left(\begin{smallmatrix}
        {\rm A} & {\rm B} & {\rm C} & {\rm D} \\ i_1 & i_2 & i_3 & i_4 \\ A & B & C & D
    \end{smallmatrix}\right)\sim(X\cdot X)^6(T\cdot T)^6.
\end{align*}
Longer contractions such as $x^{j_1}_{\rm AE}x_{j_2}^{\rm EF}x^{j_3}_{\rm FB}x^{j_4}_{\rm CD}\epsilon^{ABCD}$ are not independent and can be written as linear combinations of $(X\cdot X)\mathcal I^{x\epsilon}$. After taking into account the (anti)symmetry of various indices, we obtain a complete classification of seed R-factors, which we record in the ancillary file \texttt{possible\_Rfactors.m}. The full list of R-factors needed for the bootstrap is then obtained by permuting the $i,j,k$ labels with $S_5$ and adjusting weights so that each R-factor has conformal weight $(X\cdot X)^5(T\cdot T)^7$, $(X\cdot X)^6(T\cdot T)^6$, or $(X\cdot X)^7(T\cdot T)^5$. Note that these R-factors are not linearly independent; for example, $\mathcal I^{\epsilon\epsilon}(1234)=\mathcal I^{\epsilon\epsilon}(1235)$.

\end{appendix}

\end{document}